\documentclass[twocolumn,
showpacs,preprintnumbers,
showkeys,
amsmath,amssymb,
prl,
superscriptaddress,aps,longbibliography
]{revtex4-2}
\usepackage{amsmath}
\usepackage{amsfonts}
\usepackage{amssymb}
\usepackage{changes}
\usepackage{placeins}

\newcommand{\Rnl}{\ensuremath{R_\mathrm{nl}}}
\newcommand{\Rup}{\ensuremath{R_\mathrm{nl}^\uparrow}}
\newcommand{\Rdo}{\ensuremath{R_\mathrm{nl}^\downarrow}}
\newcommand{\Ravg}{\ensuremath{R_\mathrm{avg}}}
\newcommand{\Rscc}{\ensuremath{R_\mathrm{SCC}}}
\newcommand{\Rsccanti}{\ensuremath{R_\mathrm{SCC}^\mathrm{anti}}}
\newcommand{\Rsccsym}{\ensuremath{R_\mathrm{SCC}^\mathrm{sym}}}
\newcommand{\Rsq}{\ensuremath{R_\mathrm{sq}}}

\newcommand{\DRnl}{\ensuremath{\Delta R_\mathrm{nl}}}

\newcommand{\Vbg}{\ensuremath{V_\mathrm{bg}}}

\newcommand{\LsNbSe}[1]{\ensuremath{\lambda_s^\mathrm{NbSe_#1}}}

\newcommand{\musz}{\ensuremath{\mu_{sz}}}
\newcommand{\js}{\ensuremath{{j_s}}}
\newcommand{\jc}{\ensuremath{{j_c}}}
\newcommand{\ME}[1]{\ensuremath{M_\mathrm{#1}}}
\newcommand{\RNbSe}[1]{\ensuremath{\rho_\mathrm{NbSe_#1}}}
\newcommand{\SHAijk}{\ensuremath{\theta_{ij}^{k}}}
\newcommand{\SHAijkmin}{\ensuremath{(\theta_{ij}^{k})_\mathrm{min}}}
\newcommand{\SHAijkminabs}{\ensuremath{(|\theta_{ij}^{k}|)_\mathrm{min}}}

\begin{document}

\author{Josep Ingla-Ayn\'es}
\email{j.ingla@nanogune.eu}
\affiliation{CIC nanoGUNE BRTA, 20018 Donostia-San Sebastian, Basque Country, Spain}
\affiliation{Current address: Kavli Institute of Nanoscience, Delft University of Technology, Lorentzweg 1, Delft 2628 CJ, The
Netherlands}
\author{Inge Groen}
\affiliation{CIC nanoGUNE BRTA, 20018 Donostia-San Sebastian, Basque Country, Spain}
\author{Franz Herling}
\affiliation{CIC nanoGUNE BRTA, 20018 Donostia-San Sebastian, Basque Country, Spain}
\author{Nerea Ontoso}
\affiliation{CIC nanoGUNE BRTA, 20018 Donostia-San Sebastian, Basque Country, Spain}
\author{C. K. Safeer}
\affiliation{CIC nanoGUNE BRTA, 20018 Donostia-San Sebastian, Basque Country, Spain}
\author{Fernando de Juan}
\affiliation{Donostia International Physics Center (DIPC), 20018 Donostia-San Sebastian, Basque Country, Spain}
\affiliation{IKERBASQUE, Basque Foundation for Science, 48013 Bilbao, Basque Country, Spain}
\author{Luis E. Hueso}
\affiliation{CIC nanoGUNE BRTA, 20018 Donostia-San Sebastian, Basque Country, Spain}
\affiliation{IKERBASQUE, Basque Foundation for Science, 48013 Bilbao, Basque Country, Spain}
\author{Marco Gobbi}
\affiliation{CIC nanoGUNE BRTA, 20018 Donostia-San Sebastian, Basque Country, Spain}
\affiliation{IKERBASQUE, Basque Foundation for Science, 48013 Bilbao, Basque Country, Spain}
\author{F\`elix Casanova}
\email{f.casanova@nanogune.eu}
\affiliation{CIC nanoGUNE BRTA, 20018 Donostia-San Sebastian, Basque Country, Spain}
\affiliation{IKERBASQUE, Basque Foundation for Science, 48013 Bilbao, Basque Country, Spain}
\date{\today}
\title{Omnidirectional spin-to-charge conversion in
graphene/NbSe$_2$ van der Waals heterostructures}

\keywords{spin-to-charge conversion, van der Waals heterostructures,  symmetry}

\begin{abstract}
The conversion of spin currents polarized in different directions into charge currents is a keystone for novel spintronic devices. Van der Waals heterostructures with tailored symmetry are a very appealing platform for such a goal. 
Here, by performing nonlocal spin precession experiments, we demonstrate the spin-to-charge conversion (SCC) of spins oriented in all three directions ($x$, $y$, and $z$).
By analyzing the magnitude and temperature dependence of the signal in different configurations, we argue that the different SCC components measured are likely due to spin-orbit proximity and broken symmetry at the twisted graphene/NbSe$_2$ interface. Such efficient omnidirectional SCC opens the door to the use of new architectures in spintronic devices, from spin-orbit torques that can switch any magnetization to the magnetic state readout of magnetic elements pointing in any direction.
\end{abstract}
\maketitle
\section{Introduction}
Efficient spin-to-charge conversion (SCC) is a crucial ingredient for spintronics and has been widely studied over the last decade \cite{sinova2015}. In conventional materials with high symmetry, a charge current density (\jc{}) is converted into a spin current density (\js{}$\propto\jc{}\times s$, where $s$ is the spin polarization direction) that is perpendicular to \jc{} and $s$ via the spin Hall effect (SHE). In two-dimensional systems without structural inversion symmetry, the Edelstein effect (EE) emerges. In this case, an in-plane \jc{} leads to a perpendicularly polarized spin density ($n_s$), also in the device plane \cite{manchon2015}. Both effects obey reciprocity, implying that the inverse transformations (of \js{} and $n_s$ into \jc{}) also occur with the same efficiency in the linear response regime \cite{sinova2015}. While the SHE has been used to switch \cite{miron2011, liu2012} and probe \cite{manipatruni2019, pham2020} the magnetization of adjacent ferromagnets, the integration of spintronic devices into logic circuits still requires the introduction of materials with better SCC efficiency. Furthermore, the search for new spin manipulation approaches requires versatile materials that allow SCC of spins polarized along different directions, a feature that is not achievable in conventional materials. 
\begin{figure}[tb]
	\centering
		\includegraphics[width=0.49\textwidth]{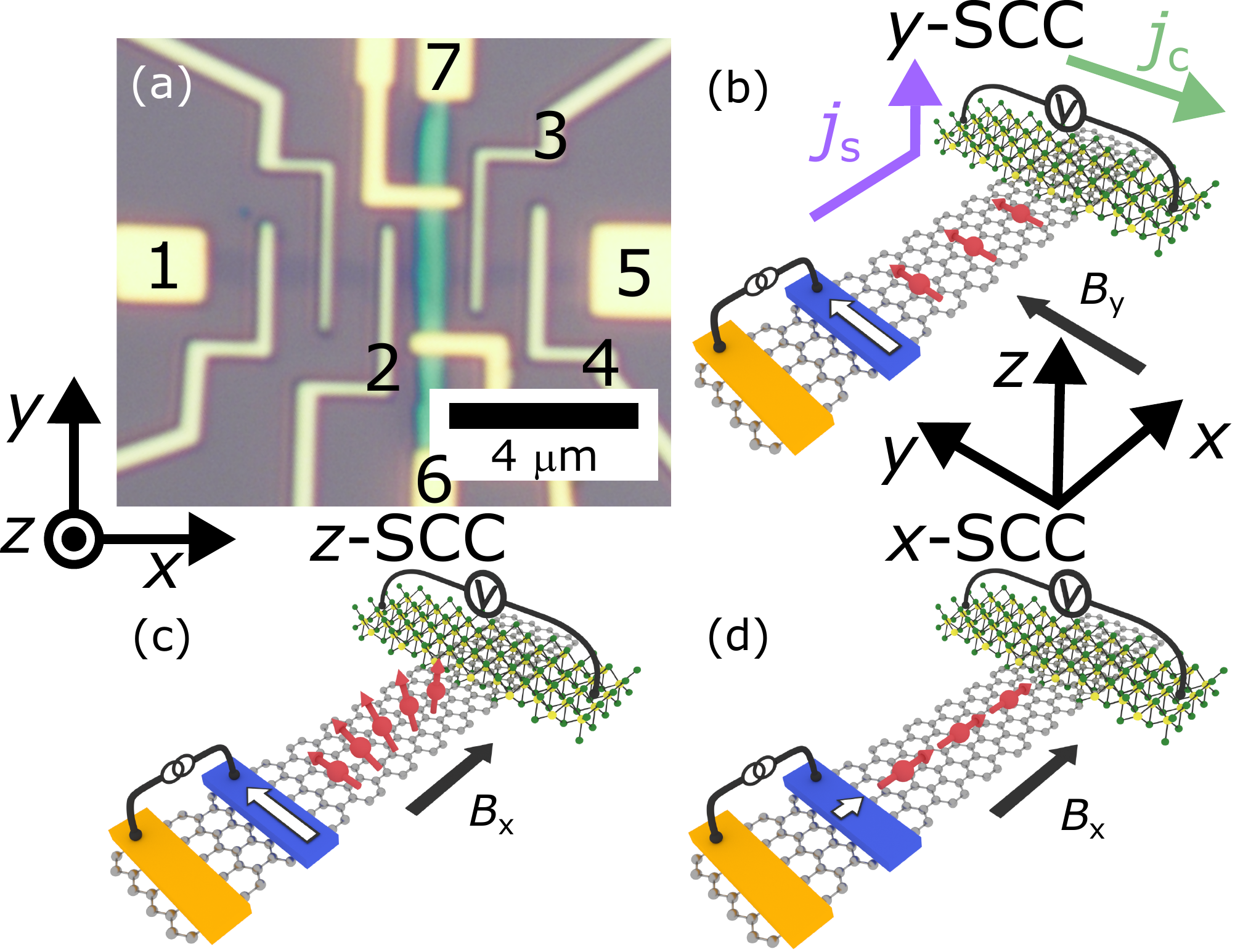}
	\caption{Device geometry implemented to detect the omnidirectional spin-to-charge conversion. (a) Optical microscope image of sample 1. The green vertical stripe is the NbSe$_2$ flake and is contacted by four Ti/Au contacts. The graphene is the dark horizontal stripe and is contacted by TiO$_x$/Co contacts (vertically oriented) and by two bigger Ti/Au contacts at the ends. (b-d) Sketches of the SCC measurement configurations used here. In b, the magnetic field is swept along the easy axis of the Co electrode, leading to magnetization switching, that reverses the sign of the $y$-SCC signal. The spin current density (\js{}) and the converted charge current density (\jc{}) directions are also shown. Note that \js{} can also have an $x$-component in the NbSe$_2$ and proximitized graphene regions. In c, a small magnetic field applied along $x$ leads to out-of-plane spin precession, enabling the measurement of $z$-SCC. In d, by increasing the magnetic field further, the Co magnetization is saturated, leading to the measurement of $x$-SCC.} 
	\label{Figure1}
\end{figure}

\begin{figure*}
	\centering
		\includegraphics[width=0.7\textwidth]{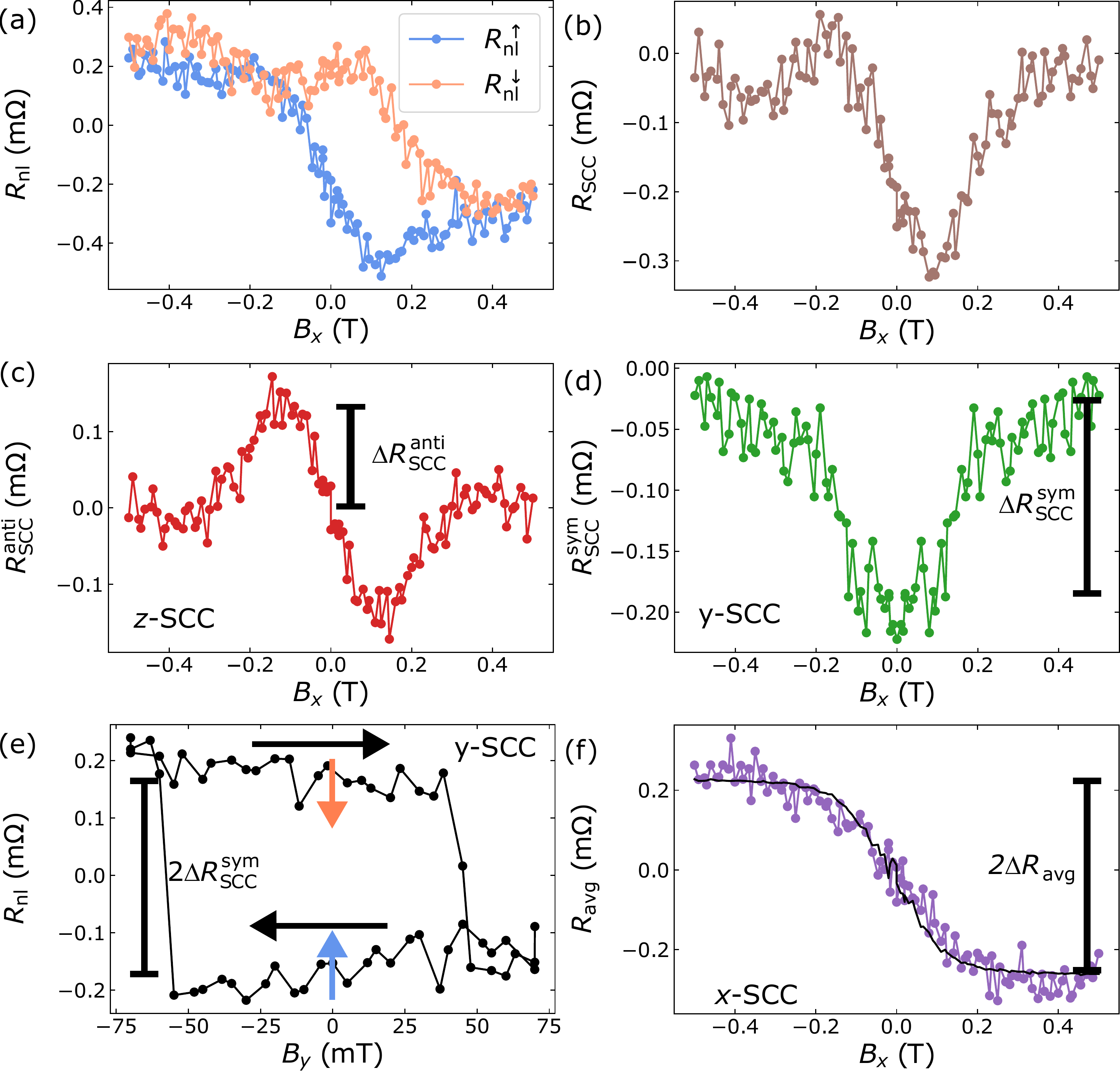}
	\caption{Nonlocal spin-to-charge conversion at 100~K and \Vbg{}=50~V. (a) \Rnl{} as a function of $B_x$ measured at the $\uparrow$ and $\downarrow$ magnetization configurations. (b) \Rscc{}$=(\Rup{}-\Rdo{})/2$ extracted from panel a. (c) Antisymmetric component of \Rscc{} vs $B_x$ corresponding to $z$-SCC with its associated amplitude $\Delta$\Rsccanti{}. (d) Symmetric component of \Rscc{} vs $B_x$ corresponding to $y$-SCC with its associated amplitude $\Delta$\Rsccsym{}. (e) \Rnl{} as a function of $B_y$. The jumps in \Rnl{} correspond to switches of the injector magnetization originated from $y$-SCC. (f) $R_\mathrm{avg}=(\Rup{}+\Rdo{})/2$ extracted from panel a with its associated amplitude $\Delta$\Ravg{}. The black line corresponds to the magnetization behaviour extracted from the reference Hanle precession data. An offset of 1.85~m$\Omega$ has been subtracted from panels a and e.}
	\label{Figure2}
\end{figure*}
In this context, layered materials emerge as a versatile platform for efficient SCC \cite{sierra2021,safeer2019,ghiasi2019,benitez2020,li2020,khokhriakov2020,herling2020,safeer2020, 1zhao2020,hoque2020,kovacs2020,galceran2021, song2020} that induces large spin-orbit torques (SOTs) in different directions \cite{macneill2017, guimaraes2018} that can switch the magnetization of adjacent ferromagnets \cite{shi2019,liu2020,hidding2020}. In particular, additional SCC components emerge in layered materials with reduced symmetry, such as 1T'-MoTe$_2$ \cite{1stiehl2019, vila2021} and T$_\mathrm{d}$-WTe$_2$ \cite{macneill2017}. In van der Waals heterostructures where these materials are combined with graphene, an additional SCC component has also been observed with $s$ parallel to the generated \jc{} \cite{ontoso2019,zhao2020}. Because this component is incompatible with the bulk symmetries of MoTe$_2$ and WTe$_2$, its origin remains elusive \cite{ontoso2019}. 

Additionally, SOTs measured in NbSe$_2$/Py bilayers have also shown the presence of an additional component not allowed by symmetry, that is sample-dependent and induced by out-of-plane spins parallel to $j_s$\cite{guimaraes2018}. This SCC component is not allowed due to the two mirror symmetries of bulk 2H-NbSe$_2$. These symmetries only enable the conversion of spins which are perpendicular to both \js{} and \jc{} via the SHE \cite{culcer2007,wimmer2015,seemann2015,roy2021}. Accordingly, new experiments combining graphene and van der Waals materials with high spin-orbit coupling (SOC) are required to understand the origin of the unexpected components. Furthermore, efficient conversions are expected from first-principles calculations, that have extracted spin-orbit proximities up to 40~meV in graphene/NbSe$_2$ heterostructures \cite{gani2020}, further motivating these experiments. Finally, even though unconventional SCC components have been observed, the achievement of omnidirectional SCC in a single device has not been realized yet.

Here, we perform nonlocal spin precession experiments to investigate SCC in graphene/NbSe$_2$ van der Waals heterostructures (Fig.~\ref{Figure1}a). Our experiments show that all three spin directions ($x$, $y$, and $z$) are converted into a charge current simultaneously while keeping the $j_c$ direction fixed (see Fig.~\ref{Figure1}b) and constitute the first realization of omnidirectional SCC. Quantitative data analysis indicates that the effects originate from at least three different SCC phenomena stemming from the SOC of NbSe$_2$, its proximity with graphene and the broken symmetry at the twisted graphene/NbSe$_2$ interface.

\section{Experimental details}
The graphene and NbSe$_2$ flakes were prepared using the conventional mechanical exfoliation technique \cite{novoselov2004} from highly oriented pyrolytic graphite and bulk NbSe$_2$ provided by HQ Graphene. The heterostructure was prepared using the PDMS-based viscoelastic transfer technique \cite{castellanos2014} in an inert atmosphere. The Ti (5~nm)/Au (120~nm) and spin-polarized TiO$_x$ (0.3~nm)/Co (35~nm) electrodes were defined using e-beam lithography and deposited using e-beam and thermal evaporation (see Supplementary Information section S1). An optical microscope image of sample 1 after fabrication is shown in Fig.~\ref{Figure1}a, where the used electrodes are numbered. The results shown here are obtained from sample 1, see the Supplementary Information section~S9 for sample 2. To optimize and keep the spin transport properties of the graphene channel nearly constant with the temperature, we have tuned the carrier density far from the charge neutrality point using a backgate voltage \cite{novoselov2004}. No SCC signals have been measured near the charge neutrality point (see Supplementary Information section~S2, S3, S6 and S9 for details on the \Vbg{} dependence of charge and spin transport in graphene and SCC in sample~2). Additionally, to minimize background effects and exclude even harmonics from our measurements, we have used the DC reversal technique with an applied current of 60~$\mu$A.
\section{Results}
To measure SCC, we use the nonlocal measurement technique, that avoids spurious effects related to local techniques such as the Oersted fields present in SOT experiments \cite{stiehl2019} and the voltages induced by stray fields measured in potentiometric measurements \cite{devries2015,li2016}. By applying a charge current ($I$) between electrodes 3 and 5, we inject a spin-polarized current in the graphene channel under electrode 3, leading to a pure spin current that diffuses to the NbSe$_2$-covered region, where it can get converted into a \jc{} via SCC in the proximitized graphene and/or absorbed by the NbSe$_2$ flake in which the SCC can subsequently occur. The \jc{} induced by SCC is detected as an open circuit voltage $V$ between the non-magnetic contacts 7 and 6, giving rise to a nonlocal signal $\Rnl{}=V/I$. Due to shape anisotropy, the easy axis of our Co electrodes is along $y$. Hence, the application of magnetic field ($B$) of more than 50~mT along $\pm y$ leads to the alignment of the electrode magnetization (Fig.~\ref{Figure1}b). By preparing the magnetization of electrode 3 (\ME3{}) along $+y\equiv\uparrow$ or along $-y\equiv\downarrow$ and applying $B$ along $\pm x$ ($B_x$, see Figs.~\ref{Figure1}c and \ref{Figure1}d), we obtain the nonlocal signals \Rup{} and \Rdo{}, respectively, which are shown in Fig.~\ref{Figure2}a. To separate between the different SCC components, we define \Rscc{}$=(\Rup{}-\Rdo{})/2$ (Fig.~\ref{Figure2}b), which contains the $y$- and $z$-SCC signals, and \Ravg{}$=(\Rup{}+\Rdo{})/2$ (Fig.~\ref{Figure2}f), which corresponds to $x$-SCC. Here, to simplify our notation, we refer to each SCC component as $x$, $y$ or $z$-SCC depending only on the spin direction.

First, we focus on the $y$- and $z$-SCC components. 
When $B_x$ is applied, the injected spins start to precess in the $y-z$ plane, leading to a net out-of-plane spin accumulation (\musz{}, see Fig.~\ref{Figure1}c). Because reversing the sign of $B_x$ leads to opposite \musz{}, the \Rnl{} expected from the $z$-SCC is antisymmetric with respect to $B_x$. Additionally, switching (the $y$ component of) \ME3{} also leads to a sign change of the signal.
Accordingly, to extract the $z$-SCC component, we have calculated the antisymmetric component of \Rscc{} (\Rsccanti{}) as a function of $B_x$. The results from Fig.~\ref{Figure2}c confirm that $z$-spins are converted in our system, giving rise to a maximum \Rsccanti{} of 0.13$\pm$0.02~m$\Omega$. We note that \Rscc{} also contains a clear symmetric component with $B_x$ (\Rsccsym{}). This component, which is shown in Fig.~\ref{Figure2}d, might correspond to a conventional Hanle spin precession measurement, that is induced by a $y$-SCC component. Since the detected spins are parallel to the injected ones, the resulting \Rscc{} decreases as the spins precess towards $\pm z$ and the $y$-spin-projection decreases symmetrically with $B_x$. We realize that $B_x$ also induces the pulling of \ME3{} along $x$, leading to a decrease of the signal as $|B_x|$ increases. Because of the short spin lifetime of our graphene channel (see Supplementary Information section~S3), we find that the lineshape is actually dominated by contact pulling, and not by spin precession. We finally note that the apparently flat trend of Fig.~\ref{Figure2}d near $B_x=0$, that is not expected from Hanle spin precession, is caused by the symmetrization of the noise in \Rscc{}.
We note that the $y$-SCC component is not expected from 2H-NbSe$_2$ and, to confirm that this SCC component is indeed present in our sample, we apply a magnetic field along $y$ ($B_y$) to control the electrode magnetization. When \ME3{} switches, the $y$-SCC signal reverses sign, allowing us to extract the signal magnitude (Fig.~\ref{Figure1}b). In Fig.~\ref{Figure2}e, we show the $B_y$-dependence of \Rnl{}. When sweeping $B_y$ from $-75$ to $+75$~mT, we observe a clear jump in \Rnl{} for $B_y\equiv B_y^{sw}=50$~mT, which is caused by the switch of \ME3{}. Additionally, sweeping $B_y$ from $+75$ to $-75$~mT leads to a jump in \Rnl{} at $B_y\approx-B_y^{sw}$, as expected for the switching of \ME3{} in the opposite direction.
Note that, because \Rscc{}$=(\Rup{}-\Rdo{})/2$, the spin signal in Fig.~\ref{Figure2}d should be approximately one half of the $0.34\pm0.03$~m$\Omega$ extracted from Fig.~\ref{Figure2}e. In agreement, we observe that the signal in Fig.~\ref{Figure2}d is of 0.16$\pm$0.02~m$\Omega$. Both signals are determined for $B=0$.

As $B_x$ increases, \ME3{} gets pulled towards $x$, leading to the injection of $x$-spins (Fig.~\ref{Figure1}d). Unlike the previous spin injection components, $x$-SCC does not depend on the initial orientation of \ME3{} in the $y$ axis. Hence, we isolate $x$-SCC by determining \Ravg{}. As shown in Fig.~\ref{Figure2}f, purple dots, the signal is antisymmetric and saturates at $\Ravg{}=0.27\pm0.06$~m$\Omega$ for $|B_x|> 0.2$~T, as expected from the contact magnetization behaviour \cite{stoner1948}. To compare \Ravg{} with the $x$-component of \ME3{} ($\ME3{}\sin(\theta_M)$, where $\theta_M$ is the magnetization angle with respect to the easy axis) extracted from spin precession in the pristine graphene region (see Supplementary Information section~S3), we have re-scaled $\sin(\theta_M)$ and plotted it as a black line in Fig.~\ref{Figure2}f. The overlap between both curves confirms that \Ravg{} follows $\sin(\theta_M)$. However, the conventional Hall effect in the graphene channel induced by the stray fields from the ferromagnetic spin injector can also lead to similar signals \cite{safeer2021, ghiasi2019}. To confirm that \Ravg{} is induced by SCC, we have measured \Rscc{} in sample 2 as a function of an out-of-plane magnetic field to induce in-plane spin precession, an unequivocal proof for spin transport (see Supplementary Information section~S9).

Finally, to compare between the different signals, in Fig.~\ref{Figure2} we define the signal amplitudes $\Delta$\Ravg{}, $\Delta$\Rsccsym{} and $\Delta$\Rsccanti{} as the semi-difference between the maximum and minimum signal vs $B_x$.
\begin{figure*}[tb!]
\centering
		\includegraphics[width=\textwidth]{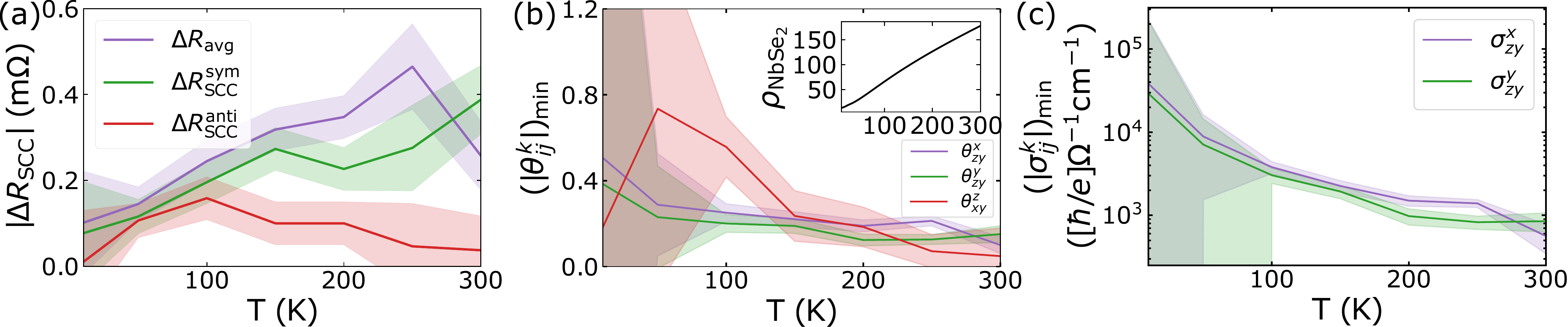}
	\caption{Temperature dependence of (a) the spin-to-charge conversion signal amplitude ($|\Delta R_\mathrm{SCC}|$) at \Vbg{}$=50$~V and the lower bound of (b) the spin Hall angle ($|\theta_{ij}^k|$), 
	and (c) the spin Hall conductivity ($|\sigma_{ij}^k|$). The inset of panel b shows the resistivity of NbSe$_2$ in $\mu\Omega$cm as a function of temperature. The shaded areas correspond to the experimental uncertainty ranges.}
	\label{Figure4}
\end{figure*}

To further understand the measured SCC, we study the temperature ($T$) dependence of $\Delta$\Ravg{}, $\Delta$\Rsccsym{} and $\Delta$\Rsccanti{}, corresponding to $x$, $y$ and $z$-SCC, respectively. In Fig.~\ref{Figure4}a, we observe that $\Delta$\Ravg{} and $\Delta$\Rsccsym{} increase with $T$ (with the exception of $\Delta$\Ravg{} at 300~K), in contrast with $\Delta$\Rsccanti{}, that shows a maximum at 100~K and decreases for higher $T$ (see Supplementary Information section S4 for the complete dataset and S5 for the reciprocity experiments that confirm that our work is in the linear response regime).
\section{Discussion}
After showing that $x$-, $y$-, and $z$-SCC occur simultaneously by spin precession measurements and how the resulting signals evolve with $T$, now we proceed to discuss the origin of the different SCC components. 

For this purpose, and because shunting prevents the accurate determination of the relevant transport parameters of proximitized graphene, we estimate the lower bound of the absolute value of the $x$-, $y$-, and $z$-SCC-associated spin Hall angles (\SHAijkminabs{}, where $i$, $j$, and $k$ are the directions of \js{}, \jc{} and $s$, respectively) assuming that the SCC occurs in the NbSe$_2$ flake (see Supplementary Information section~S8 for details). 
The results from our analysis are shown in Fig.~\ref{Figure4}b as a function of $T$. We observe that the error range (shaded areas) increases dramatically upon cooling below 100~K. This effect is caused by the decrease in \RNbSe2{} (that is assumed to be isotropic) shown in the inset. The decrease in the device resistance with decreasing $T$ leads to much smaller signals while keeping \SHAijkminabs{} constant, decreasing the sensitivity of our measurement below 100~K. We stress that the resulting \SHAijkminabs{} values are only valid under the assumption that the conversion occurs in the NbSe$_2$ flake. If the SCC occurs in the proximitized graphene, lower efficiencies are to be expected.

\begin{figure*}[tb!]
	\centering
		\includegraphics[width=\textwidth]{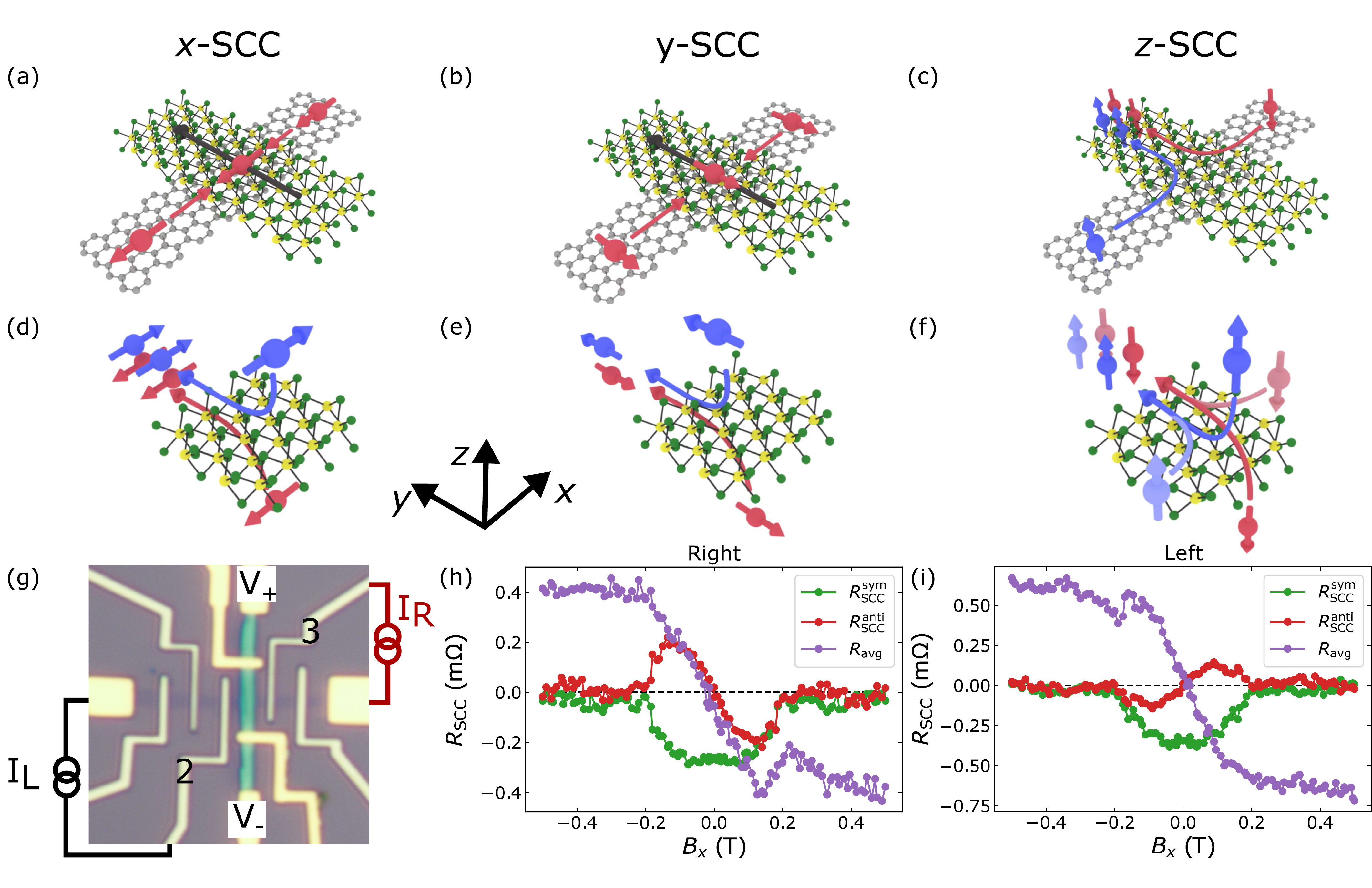}
	\caption{(a)-(c) Proximity-induced SCC in graphene/NbSe$_2$ heterostructures. Conventional (a) and unconventional (b) inverse EE in proximitized graphene lead to the conversion of $x$- and $y$-spins, respectively, into a $y$-charge current. (c) Inverse SHE in proximitized graphene leads to $z$-SCC. (d)-(f) Possible SHE components originating in the NbSe$_2$ flake. (d) A spin current polarized along $x$ and propagating along $z$ gives rise to a charge current along $y$ via the conventional inverse SHE. (e) Under shear strain, a spin current polarized along $y$ and propagating along $z$ can give rise to a $y$-charge current in NbSe$_2$. (f) A spin current polarized along $z$ and propagating along $z$ and $x$ can give rise to a $y$-charge current via unconventional and conventional (inverse) SHE, respectively. 
	(g) SCC is measured at 100~K and \Vbg{}$=70$~V using a spin current source placed at the right ($I_\mathrm{R}$) (h) and left ($I_\mathrm{L}$) (i) sides of the NbSe$_2$ flake. An offset of 2.3 and 1.5~m$\Omega$ has been subtracted from \Ravg{} in panels (h) and (i), respectively.}
	\label{Figure3}
\end{figure*}

First, we discuss the $z$-SCC component. It can be caused by the inverse SHE in proximitized graphene \cite{safeer2019, benitez2020, herling2020,safeer2020} (Fig.~\ref{Figure3}c), by the conventional inverse SHE in NbSe$_2$ due to a spin current polarized along $z$ and diffusing along $x$ (Fig.~\ref{Figure3}f light red and blue spins), or due to the unconventional out-of-plane SCC component in NbSe$_2$ (Fig.~\ref{Figure3}f red and blue spins) reported in Ref.~\cite{guimaraes2018}. The $z$-SCC origin can be discerned by changing the sign of $j_s$ in the graphene channel (along $x$-axis). If the SCC occurs in the proximitized graphene channel or by conventional SHE in NbSe$_2$, because the relevant \js{} flows along $x$, \Rscc{} must change sign with $j_s$. In contrast, if the SCC occurs via unconventional out-of-plane SCC in the NbSe$_2$ flake, because the relevant $j_s$ points along $z$ in the NbSe$_2$, the sign of \Rscc{} should not change when changing the in-plane \js{} direction. To discern between these effects, we compare between \Rnl{} obtained using electrodes 3 and 2 to inject spins (Fig.~\ref{Figure3}g). The results are shown in Figs.~\ref{Figure3}h and \ref{Figure3}i, respectively. We observe that \Rsccsym{} and \Ravg{} do not change sign when changing the spin injector (i.e. the in-plane \js{} direction) but \Rsccanti{} does, indicating that the $z$-SCC is induced by the inverse SHE in the proximitized graphene channel or the NbSe$_2$ flake. Now we need to discern between both possibilities.
Looking at the spin Hall angle, we note that $|\theta_{xy}^z|$ is higher than 50\% at 100~K. The quantification performed here assumes that \js{} is absorbed all along $z$ in the NbSe$_2$. Because the $x$-component of \js{} in NbSe$_2$ is smaller than the total \js{}, the red line in Fig.~\ref{Figure4}b provides an even lower bound to $|\theta_{xy}^z|$ and the true spin Hall angle is expected to be higher. In this context, since 50\% is already significantly larger than observed previously in NbSe$_2$ devices\cite{guimaraes2018}, it is safe to assume that the $z$-SCC component originates from the proximitized graphene layer instead of the NbSe$_2$. Due to shunting by the NbSe$_2$, we cannot obtain the charge and spin transport parameters of the proximitized graphene region. Since these parameters are required to quantify the proximity-induced SCC efficiency, the subsequent quantitative analysis (Fig.~\ref{Figure4}c) assuming that the SCC occurs in the NbSe$_2$ channel is not performed for the $z$-SCC component.

 Next, we address the in-plane SCC components. $x$-SCC can be induced by both the conventional EE in the proximitized graphene (Fig.~\ref{Figure3}a) and the conventional SHE in NbSe$_2$ (Fig.~\ref{Figure3}d). The lack of a sign reversal of the $x$-SCC component when reversing the spin current direction (Fig.~\ref{Figure3}g-i) is consistent with both possibilities. As a consequence, we cannot discern between them by symmetry considerations. In bulk NbSe$_2$, the $y$-SCC component should in principle be forbidden by symmetry \cite{guimaraes2018}. However, at the first NbSe$_2$ layer, which has C$_{3v}$ symmetry, there is an allowed SHE component that would convert $y$-spins propagating along $x$ into a charge current along $y$ \cite{culcer2007, wimmer2015,seemann2015,roy2021}. However, the lack of a sign reversal of the $y$-SCC component when reversing the spin current direction (Fig.~\ref{Figure3}g-i) rules out this possibility. Additionally, we note that for any twist angle between the crystal mirrors of the NbSe$_2$ and graphene flakes that is not a multiple of 30 degrees, the graphene and NbSe$_2$ vertical mirrors are not aligned, so that neither can be a mirror for the whole structure, resulting in a C$_3$ point group that does not have any mirror symmetry, enabling $y$-SCC via unconventional EE in the proximitized graphene \cite{li2019,david2019} as shown in Fig.~\ref{Figure3}b (see Supplementary Information section S11 for a detailed symmetry discussion). Furthermore, recent first-principles calculations on twisted graphene/transition metal dichalcogenide heterostructures \cite{naimer2021, pezo2021} have shown that the radial component of the in-plane SOC can have a similar magnitude as the conventional Rashba SOC component \cite{naimer2021} depending on the electric field and twist angle, indicating that the measured component is likely to arise from the EE in the graphene channel. Finally, shear strain could also break enough symmetries in the NbSe$_2$ flake, enabling a SCC component where $y$-polarized spins flow in the $z$ direction in NbSe$_2$ (Fig.~\ref{Figure3}e) and propagate into the graphene channel\cite{ontoso2019}. The lack of a sign change of  \Rsccsym{} when reversing the spin current direction (Fig.~\ref{Figure3}g-i) is compatible with the two mentioned mechanisms.
 
 To gain insight into the origin of the in-plane SCC components, we discuss the $\SHAijkminabs{}$ shown in Fig.~\ref{Figure4}b that assumes that the SCC occurs in the NbSe$_2$ flake. 
 We observe that, at 100~K, in-plane $\SHAijkminabs{}$ reaches the highest value, which is 25 and 20\% for $x$- and $y$-SCC, respectively. At higher $T$, the measured signals increase, leading to minimum \SHAijkminabs{} values of 10 and 15\% at 300~K for spins polarized along $x$ and $y$, respectively.
 
 We can compare our results with SOT experiments by calculating the spin Hall conductivity $(\sigma_{ij}^k)_\mathrm{min}=\SHAijkminabs{}/\RNbSe2{}$. We find that, at 300~K, $\sigma_{zy}^x=560\pm200$ and $\sigma_{zy}^y=850\pm200$~$[\hbar/e]\Omega^{-1}$cm$^{-1}$. These values, which are already large at 300~K, increase an order of magnitude upon cooling to 100~K. Note that the extracted $\sigma_{ij}^k$ are one order of magnitude larger than the conductivities of up to 75~$[\hbar/e](\Omega$ cm$)^{-1}$ extracted from SOT experiments \cite{guimaraes2018}. 
The large difference between our results and those of Ref.~\citenum{guimaraes2018} suggests that both in-plane SCC components, that have similar magnitude, occur in the proximitized graphene layer. Note that, in this case, the injected spins can be converted without overcoming the interface resistance with the NbSe$_2$. flake (see Supplementary Information section S7 for the determination of the interface resistance). Accordingly, the conversion efficiency required to explain the measured signals is expected to decrease significantly.
Furthermore, the reproducible observation of the unconventional $y$-SCC component in graphene-based heterostructures containing different layered metals (Refs.~\citenum{ontoso2019} and \citenum{zhao2020}) suggests that breaking of the symmetry at the interface is the most plausible source for such a conversion. Since the strain has been shown to fluctuate randomly in graphene-based van der Waals heterostructures \cite{couto2014, neumann2015}, the consistent occurrence of shear strain in all the devices seems less likely than an imperfect alignment effect. It is also worth noting that, unlike the case of graphene/semiconducting transition metal dichalcogenide devices, the  samples mentioned here are not annealed at high temperatures, minimizing the probability of crystallographic alignment between the different materials \cite{wang2015}. Despite not having an annealing step, we have found that the interface resistances between the graphene and NbSe$_2$ flakes are lower than 200~$\Omega$ in the reported devices, demonstrating that the interface is transparent enough to induce proximity on the graphene flake (see Supplementary Information section S7 for details).
For all these reasons, we believe that the signal quantification in Figs.~\ref{Figure4}b and \ref{Figure4}c most likely does not reflect the actual SCC efficiency in our device. However, it shows that proximitized graphene allowed us to measure larger SCC signals than bulk NbSe$_2$ alone.

To infer whether the twist angle between the graphene and NbSe$_2$ flakes is a multiple of 30$^\circ$, an alignment that would be incompatible with our interpretation, we have assumed that the straight edges of the flakes correspond to crystallographic directions and used the optical microscope images to estimate the twist angle between the graphene and NbSe$_2$ flakes in sample~1. The result is 89$^\circ\pm0.6^\circ$ (see Supplementary Information section S10) and, even though this angle is rather small (it is only $1^\circ$ from a high-symmetry point), first-principles calculations predict that the radial component of the spin texture depends more strongly on band alignments and electric fields than on the twist angle \cite{naimer2021}. As a consequence, we believe that the small twist angle does not contradict the broken-symmetry interpretation.

Finally, we argue that the $T$-dependence of the SCC signal is consistent with our interpretation. Even though the $x$- and $y$-SCC signals increase with $T$, in contrast with the EE in graphene proximitized by a semiconductor \cite{ghiasi2019}, this different behavior can be attributed to the role of NbSe$_2$ as a shunting layer. As shown in the inset of Fig.~\ref{Figure4}b, \RNbSe2{} increases dramatically with $T$, reducing the shunting and thus increasing the SCC signal. In contrast, the $z$-SCC component decreases when increasing $T$. Theoretically, it has been predicted that the SHE in proximitized graphene depends on the intervalley scattering time \cite{garcia2017,milletari2017, garcia2018}. In this context, the metallic NbSe$_2$ flake may induce extra intervalley scattering and have a detrimental effect on the $z$-SCC signal at high $T$.

\section{Summary}
In summary, we have shown that graphene/NbSe$_2$ van der Waals heterostructures convert $x$, $y$, and $z$ spins simultaneously. 
By analyzing the magnitude, $T$-dependence and symmetry of the SCC signals, we argue that the three components are likely to occur at the NbSe$_2$-proximitized graphene.
In particular, the $z$-SCC component most likely arises due to SHE, the $x$-component due to the EE, and the $y$-SCC component due to unconventional EE at the twisted graphene/NbSe$_2$ heterostructure. A twist angle between the flakes can break all the mirror symmetries and enable the unconventional EE component. 
Our discovery of omnidirectional SCC paves the way for novel spintronic devices that use the three spin directions to realize complex operations. For instance, spins in different directions can be controlled independently via SOC-induced spin precession \cite{ingla2021} and contribute to the output signal, enabling the realization of new spin-based operations.

\section{Acknowledgments}
We acknowledge R.~Llopis and R.~Gay for technical assistance. This work is supported by the Spanish MICINN under projects RTI2018-094861-B-I00, PID2019-108153GA-I00 and the Maria de Maeztu Units of Excellence Programme (MDM-2016-0618 and CEX2020-001038-M), by the ``Valleytronics" Intel Science Technology Center, and by the European Union H2020 under the Marie Sklodowska-Curie Actions (0766025-QuESTech). J.I.-A. acknowledges postdoctoral fellowship support from the “Juan de la Cierva - Formación” program by the Spanish MICINN (Grant No. FJC2018-038688-I). N.O. thanks the Spanish MICINN for
a Ph.D. fellowship (Grant no. BES-2017-07963). C.K.S. acknowledges support from the European
Commission for a Marie Sklodowska-Curie individual fellowship (Grant No.
794982-2DSTOP). F. J. acknowledges funding from the Spanish MCI/AEI/FEDER through grant PGC2018-101988-B-C21 and from the Basque government through PIBA grant 2019-81. M.G. acknowledges support from la Caixa Foundation for a Junior Leader fellowship (Grant No. LCF/BQ/PI19/11690017).


\bibliography{bibliographySI}

\textcolor{white}{Dummy text}

\widetext
\begin{center}
\textbf{\large Supplementary information of "Omnidirectional spin-to-charge conversion in graphene/NbSe$_2$ van der Waals heterostructures"}
\end{center}
\setcounter{equation}{0}
\setcounter{figure}{0}
\setcounter{table}{0}
\setcounter{page}{1}
\makeatletter
\renewcommand{\theequation}{S\arabic{equation}}
\renewcommand{\thefigure}{S\arabic{figure}}
\newpage
\tableofcontents

\renewcommand{\figurename}{Figure}
\renewcommand{\thetable}{S\arabic{table}}
\renewcommand{\theequation}{S\arabic{equation}}
\renewcommand{\thefigure}{S\arabic{figure}}
\renewcommand\thesection{S\arabic{section}}

\newcommand{\Rpar}{\ensuremath{R_\mathrm{nl}^\mathrm{P}}}
\newcommand{\Ranti}{\ensuremath{R_\mathrm{nl}^\mathrm{AP}}}
\newcommand{\Rdown}{\ensuremath{R_\mathrm{nl}^\downarrow}}
\newcommand{\Rgr}{\ensuremath{R_\mathrm{sq}^\mathrm{gr}}}
\newcommand{\Rint}{\ensuremath{R_\mathrm{int}}}
\newcommand{\Rmeas}{\ensuremath{R_\mathrm{meas}}}

\clearpage
\section{Device fabrication}
\begin{figure}[h]
	\centering
		\includegraphics[width=0.8\textwidth]{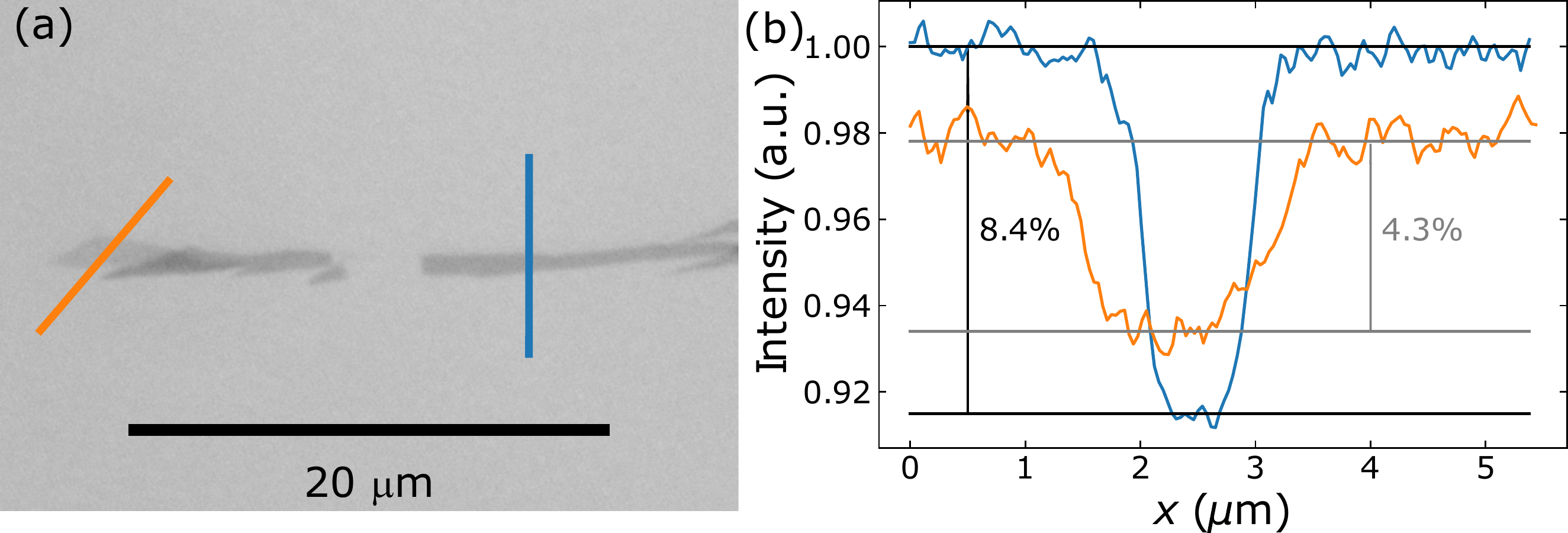}
	\caption{(a) Red channel of the optical microscope image of the bilayer graphene flake used for device 1 (right) and a monolayer graphene region (left). (b) Optical contrast of the monolayer (4.3\%) and bilayer (8.4\%) graphene flakes. The profiles are taken along the lines in panel a.}
	\label{FigureOpticalContrast}
\end{figure}
The bilayer graphene (BLG) flake was obtained by cleaving a highly oriented pyrolytic graphite crystal (provided by HQ graphene) on a Si substrate with 300 nm of thermal oxide using Nitto tape. To determine the number of layers of the exfoliated flakes, we used optical contrast. Fig.~\ref{FigureOpticalContrast} shows that the optical contrast of BLG is twice the one of monolayer graphene.

\begin{figure}[h]
	\centering
		\includegraphics[width=0.8\textwidth]{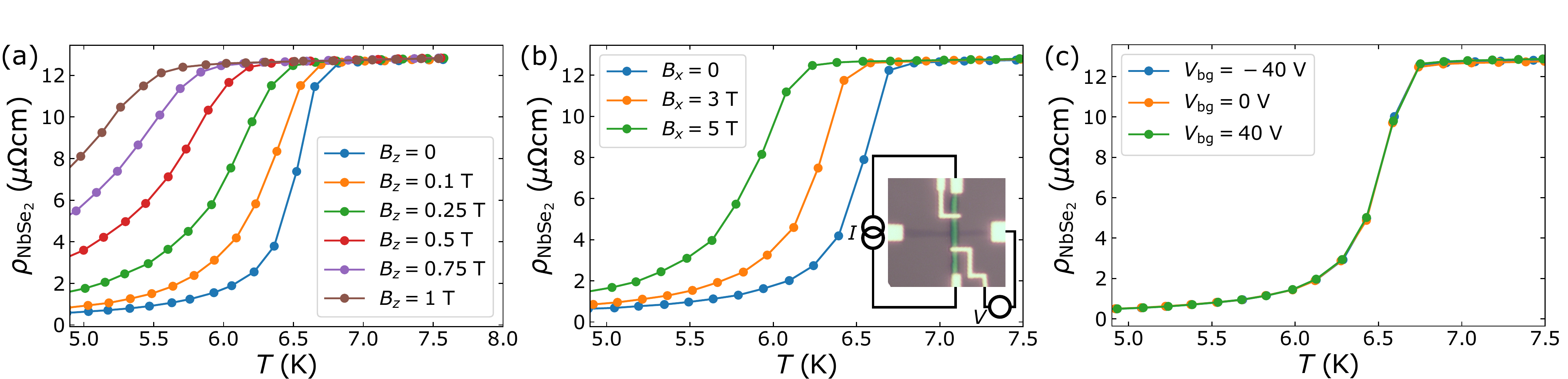}
	\caption{Temperature dependence of the NbSe$_2$ resistivity for different $B_x$ (a), $B_z$ (b) and \Vbg{} (c). The measurement circuit is shown at the inset of panel b.}
	\label{FigureSuper}
\end{figure}

The NbSe$_2$ flakes were exfoliated from a bulk NbSe$_2$ crystal (provided by HQ graphene) on PDMS  (Gelpack 4) in a glove box with an Ar atmosphere. The (16-nm-thick) NbSe$_2$ flake employed in device 1 was transferred on top of the BLG flake shown in Fig.~\ref{FigureOpticalContrast}a using the viscoelastic stamping technique \cite{castellanos2014} in the glove box. Next, the Ti (5~nm)/Au (120~nm) contacts were defined using conventional e-beam lithography and deposited by e-beam and thermal evaporation, respectively. To prevent oxidation of the NbSe$_2$ flake, we minimized the exposure of the flake to air.
At this stage, we characterized the NbSe$_2$ flake by measuring its resistivity  (\RNbSe2{}) as a function of $T$ and at different in-plane ($B_x$), out-of-plane ($B_z$) magnetic-fields, and backgate voltages (\Vbg{}) (Figure~\ref{FigureSuper}).

\begin{figure}[h]
	\centering
		\includegraphics[width=0.4\textwidth]{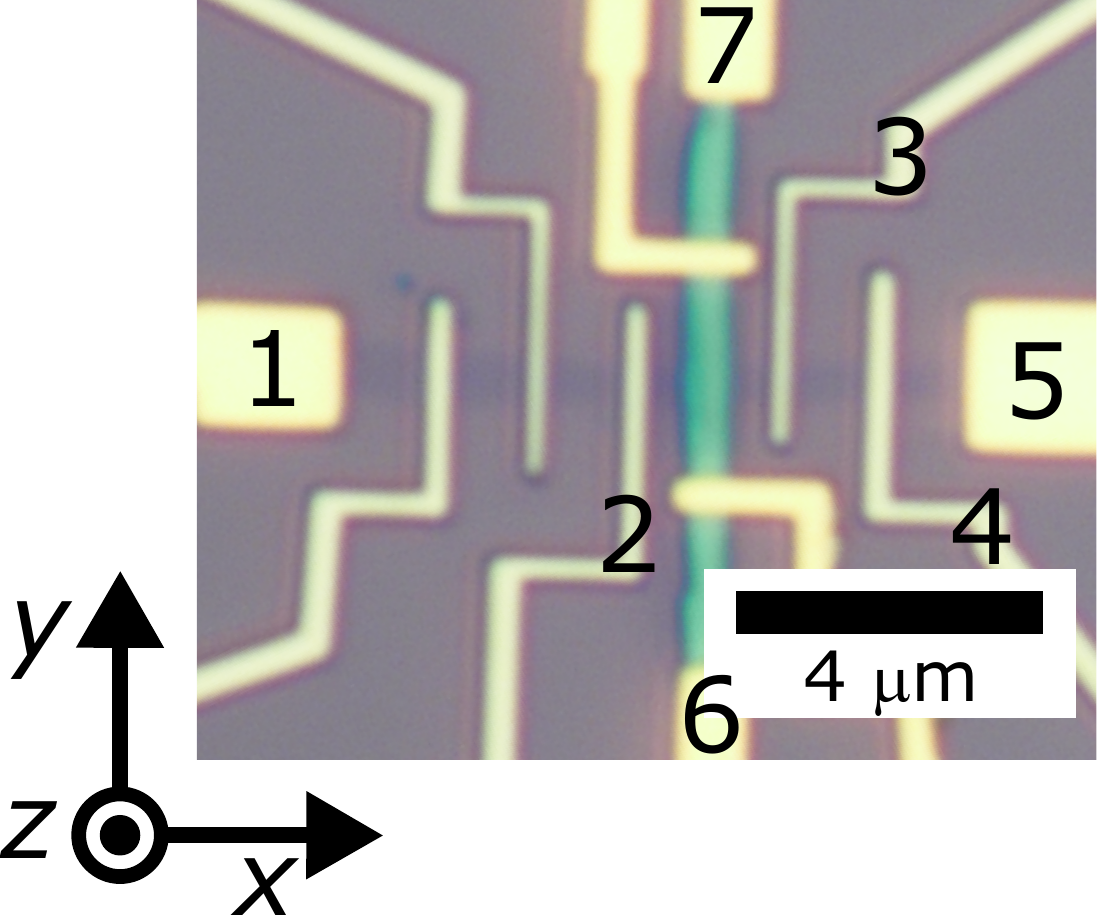}
	\caption{Optical microscope image of the device with the corresponding contact numbering. The grey horizontal stripe is the BLG flake, the green vertical stripe corresponds to the NbSe$_2$, the vertical contacts to the BLG flake are TiO$_x$/Co and the other electrodes are Ti/Au.}
	\label{FigureDeviceImage}
\end{figure}

Finally, the TiO$_x$/Co electrodes were defined using e-beam lithography and e-beam evaporation of 0.3~nm of Ti, followed by oxidation in air for 10 min, and 35~nm of Co. The ferromagnetic electrodes were capped with 5~nm of Au to prevent oxidation.
An optical microscope image of the completed device is shown in Fig.~\ref{FigureDeviceImage}.

In total, we prepared 8 devices out of which we could measure spin transport in three of them: The two shown in the report and one where one NbSe$_2$ arm broke and we could only measure passing a current between the NbSe$_2$ and graphene flakes. This device showed higher signals, but it is very hard to compare with the others as the measurement geometry is not the same, so we kept it aside. In the 5 remaining devices either the electrical contacts to NbSe$_2$ were not good  enough to measure SCC or the TiO$_x$/Co contacts were not spin polarized.
\section{Analysis of the \Vbg{} sweeps}\label{SectionGateSweep}
\subsection{Measurement of the square resistance of the pristine BLG region}
\begin{figure}[h]
	\centering
		\includegraphics[width=0.75\textwidth]{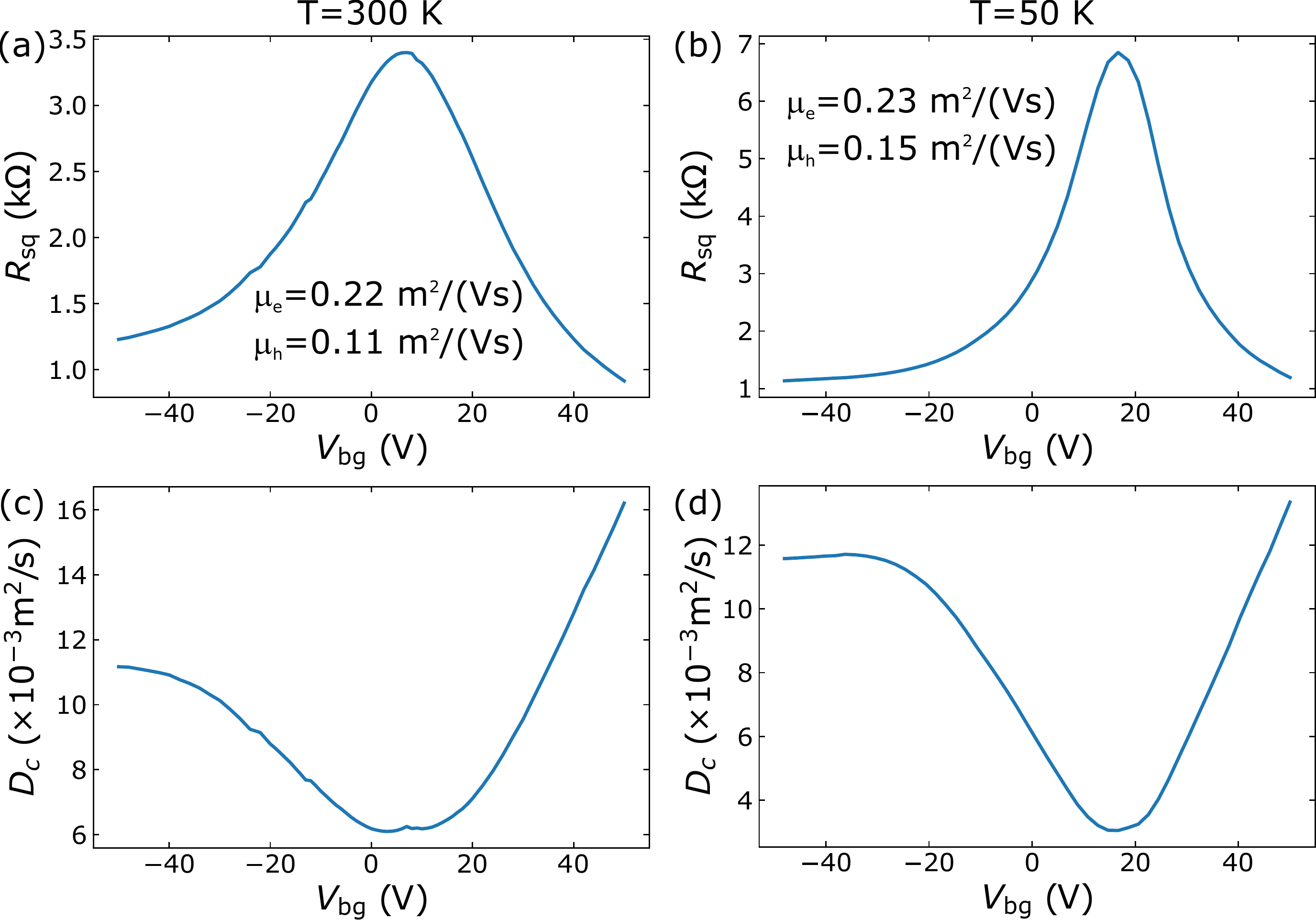}
	\caption{Square resistance \Rgr{} and charge diffusivity $D_c^\mathrm{gr}$ of the pristine BLG region as a function of \Vbg{} at 300~K [panels (a) and (c)] and at 50~K [panels (b) and (d)].}
	\label{FigureRsq}
\end{figure}

To determine the charge transport properties of the device shown in the main manuscript and Fig.~\ref{FigureDeviceImage}, we measured the channel's square resistance (\Rsq{}) as a function of the backgate voltage (\Vbg{}), that is applied to the doped Si substrate \cite{novoselov2004}. The \Vbg{} controls the carrier density ($n$) in the graphene channel via the field effect,

\begin{equation}
n=\frac{\epsilon_0\epsilon_r}{et_\mathrm{SiO_2}}(V_\mathrm{bg}-V_\mathrm{cnp}),
\label{EquationS1}
\end{equation}
where $\epsilon_{0}$ is the vacuum dielectric permittivity, $\epsilon_{r}=3.9$ is the dielectric constant of SiO$_2$, $e$ the electron charge, $t_\mathrm{SiO_2}=300$~nm is the thickness of the SiO$_2$ dielectric, and $V_\mathrm{cnp}$ the value of \Vbg{} at which the graphene reaches the charge neutrality point (CNP).
In Fig.~\ref{FigureRsq}a, we show \Rsq{} vs \Vbg{} at the pristine graphene region at 300~K obtained by measuring the voltage drop between contacts 3 and 4 ($V_{34}$) while applying a current between contacts 6 and 5 ($I_{65}=1\,\mu$A). \Rgr{} is determined using 
\begin{equation}
R_\mathrm{sq}^\mathrm{gr}=(V_{34}/I_{65})(W_\mathrm{gr}^{34}/L_{34}),
\label{EquationS2}
\end{equation}
where $W_\mathrm{gr}^{34}=1.0\,\mu$m is the average sample width between contacts 3 and 4 and $L_{34}=2.0\,\mu$m is the spacing between contacts 3 and 4 (Table~\ref{TableS1}). From Fig.~\ref{FigureRsq}a, one can observe that \Rsq{} shows a clear peak for \Vbg{}$=V_\mathrm{cnp}\approx\,$7~V, which corresponds to the CNP, implying that the sample is slightly $p$-doped at \Vbg{}$=$0~V. At 50~K we see that the position of $V_\mathrm{cnp}$ has shifted to $\approx\,$17~V.

\subsection{Determination of the charge diffusivity}
Because of the weak electron-electron interactions in graphene \cite{bandurin2016}, in samples with moderate mobility, the charge ($D_c$) and spin diffusivity ($D_s$) can be assumed to be equal \cite{maassen2011}. Hence, it is useful to obtain $D_c$ from the \Vbg{} sweeps. For this purpose, we use the Einstein relation $D_c=(e^2R_\mathrm{sq}\nu(E_F))^{-1}$, where $\nu(E_F)$ is the density of states at the Fermi level. Using the density of states of BLG, the following expression is obtained: 
\begin{equation}
D_c=\frac{\pi\hbar^2 v_{f0}^2}{R_\mathrm{sq}e^2\sqrt{\gamma_1^2+4\pi\hbar^2v_{f0}^2|n|}},
\label{EinsteinRelation}
\end{equation}

where $v_{f0}=1\times10^6$~m/s is the Fermi velocity of graphene, $\gamma_1\sim0.4$~eV is the interlayer coupling parameter between pairs of orbitals on the dimmer sites in BLG \cite{mccann2013}, and $\hbar$ is the reduced Planck constant. Using Equation~\ref{EinsteinRelation}, the measured \Rsq{}, and $n$ (obtained using Equation~\ref{EquationS1}) we obtain $D_\mathrm{c}^\mathrm{gr}$ for the pristine graphene region. These results are shown in Figs.~\ref{FigureRsq}c-d. 
Finally, to determine the charge transport quality of our device, we calculated the field-effect electron (hole) mobility ($\mu_{e(h)}$) using $\Rsq{}^{-1}=ne\mu_{e(h)}$, at $15\,\mathrm{V}<|\Vbg{}-V_\mathrm{cnp}|<40$~V. The results are shown in Fig.~\ref{FigureRsq}. We observe that the obtained mobilities are of about 2~m$^2/$(Vs), as expected for graphene on SiO$_2$ devices. Additionally, we observe that 1/\Rsq{} and $D_\mathrm{c}$ are higher at 300~K than at 50~K for all the \Vbg{} range except for $\Vbg{}\gtrsim-30$~V, where  $D_\mathrm{c}$ starts to saturate at both temperatures. We attribute this observation to the thermal broadening.
\section{Hanle precession at the pristine graphene region}

\begin{figure}[h]
	\centering
		\includegraphics[width=0.75\textwidth]{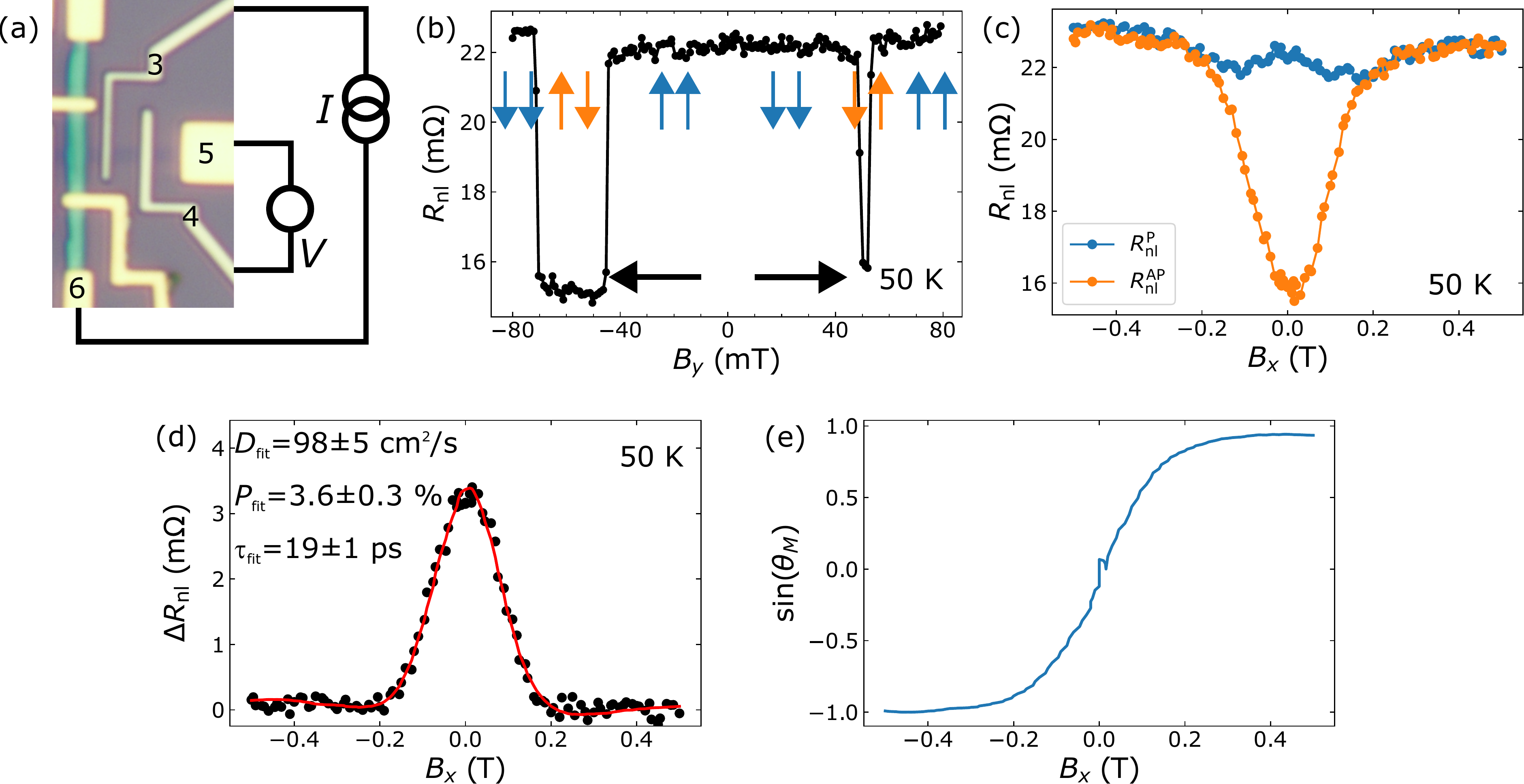}
	\caption{Spin transport in the pristine graphene region. (a) Measurement configuration. (b) Nonlocal resistance as a function of a magnetic field parallel applied along $y$. The black arrows indicate the $B_y$ sweep direction and the blue and orange arrows indicate the spin injector and detector magnetization configuration obtained assuming that contact 3 switches after 4. (c) Nonlocal resistance as a function of a magnetic field parallel applied along $x$ in the parallel and antiparallel magnetization configurations. (d) \DRnl{} vs $B_x$ obtained from panel c, together with its fit to the solution of the Bloch equations and the extracted parameters. (e) $B_x$-induced pulling of the contact magnetization. The data of panels b-e was obtained at $T=50$~K and \Vbg{}$=50$~V.}
	\label{FigureRefHanle}
\end{figure}
To extract the spin Hall angle from the measured signals, it is required to obtain the spin transport parameters of the graphene channel.
For this purpose, we have performed nonlocal spin transport experiments between contacts 3 and 4 (Fig.~\ref{FigureDeviceImage}).
 As shown in Fig.~\ref{FigureRefHanle}a, by applying a charge current ($I$) between contacts 3 and 6, a spin current is injected in the graphene channel. The spin current induces a spin accumulation that diffuses in the graphene channel and is detected by measuring the voltage ($V$) between contacts 4 and 5. When applying a magnetic field along $y$ ($B_y$) antiparallel to the contact magnetizations (Fig.~\ref{FigureRefHanle}b), the magnetizations of contacts 3 and 4 are switched. Because these electrodes have different coercivities, the switches occur at different $B_y$ values. In this context, measuring the nonlocal resistance (\Rnl{}$=V/I$) as a function of $B_y$ allows the determination of the spin signal in the parallel (\Rpar{}) and antiparallel (\Ranti{}) configurations (Fig.~\ref{FigureRefHanle}b).
We determine the spin transport properties of the pristine graphene region by performing Hanle spin precession measurements. For this purpose, we apply a magnetic field along $x$, that induces out-of-plane spin precession (in the $y-z$ plane) before pulling the contact magnetizations along $x$. In particular, the measured signal in the parallel (antiparallel) configuration can be written as:
\begin{equation}
R_\mathrm{nl}^\mathrm{P(AP)}=+(-)R_\mathrm{prec}\cos^2(\theta_M)+R_0\sin^2(\theta_M)
\end{equation}
Where $R_\mathrm{prec}$ is the spin precession signal that is the solution of the Bloch equations and includes the spin backflow at the contacts \cite{maassen2012}, $\theta_M$ is the magnetization angle with respect to its easy axis, and $R_0=R_\mathrm{prec}(B_x=0)$. 

The result of measuring \Rpar{} and \Ranti{} vs $B_x$ is shown in Fig.~\ref{FigureRefHanle}c. The spin signal \DRnl{}$=(\Rpar{}-\Ranti{})/2$ is shown in Fig.~\ref{FigureRefHanle}d, together with its fit to $R_\mathrm{prec}\cos^2(\theta_M)$, where the contact pulling is obtained using
\begin{equation}
\sin(\theta_M)=\sqrt{\frac{R_\mathrm{avg}-\min(R_\mathrm{avg})}{\max(R_\mathrm{avg}) - \min(R_\mathrm{avg})}},
\end{equation}
where $R_\mathrm{avg}=(\Rpar{}+\Ranti{})/2$. The result from such an operation is shown in Fig.~\ref{FigureRefHanle}e.

The parameters extracted from the fit are shown in Fig.~\ref{FigureRefHanle}d. We note that the extracted spin lifetime ($\tau_\mathrm{fit}=19$~ps) is significantly shorter than typical values obtained in graphene, which are in the 100~ps range. A priori, one could expect that spins are absorbed by the NbSe$_2$, leading to shorter spin lifetimes in the graphene channel. However, the extracted spin relaxation length ($\lambda_\mathrm{fit}=\sqrt{\tau_\mathrm{fit}D_\mathrm{fit}}\approx0.4\,\mu$m) is shorter than the distance between contact 3 and the nearest edge of the NbSe$_2$, which is 1~$\mu$m, implying that the NbSe$_2$ cannot be the reason for the short $\lambda_\mathrm{fit}$.
Because backflow is also included in $R_\mathrm{prec}$, and the contact resistances are 9 and 5~k$\Omega$, we conclude that the spin relaxation must occur in the graphene channel. Thus, the most likely reason for the short $\lambda_\mathrm{fit}$ is a reduction of the spin lifetime in the graphene channel under the contacts. This effect has been observed in samples with moderate contact resistances where the separation between contacts is short \cite{amamou2016}. It is the case in our sample where the contact separation is 1~$\mu$m.

The spin transport measurements in Fig.~\ref{FigureRefHanle}, which are taken at \Vbg{}$=50$~V and T$=50$~K, have also been performed at other temperatures and the results are shown in Table~\ref{TableRefHanle}. As it can be seen from there, the results are very similar for the three $T$ values. Even though $D_\mathrm{fit}$ at 300~K is significantly lower than at 100 and 50~K, we note that it gets compensated by a higher $P_\mathrm{fit}$. We attribute this to the fact that $D_\mathrm{fit}$ is determined by the \Rnl{} values at high $B_x$, which are dominated by contact pulling, the extracted values of $D_\mathrm{fit}$ may have extra uncertainties.
\begin{table}[t]
    \caption{Spin transport parameters pristine graphene at \Vbg{}$=50$~V.}
        \begin{ruledtabular}
        \begin{tabular}{c c c c}
            \renewcommand{\arraystretch}{2}
            $T$ & $D_\mathrm{fit}$ & $\tau_\mathrm{fit}$&$P_\mathrm{fit}$\\
			(K)	&(cm$^2$/s)&(ps)	&(\%)\\
            \hline
             300 & 62$\pm$7 &21$\pm$1&5$\pm$1\\
             100 & 109$\pm$7 &19$\pm$1&3.4$\pm$0.2\\
             50 & 98$\pm$5 &19$\pm$1&3.6$\pm$0.2\\
    \end{tabular}
    \end{ruledtabular}
    \label{TableRefHanle}
\end{table}

\section{Spin-to-charge conversion experiments}
\begin{figure}[h]
	\centering
		\includegraphics[width=\textwidth]{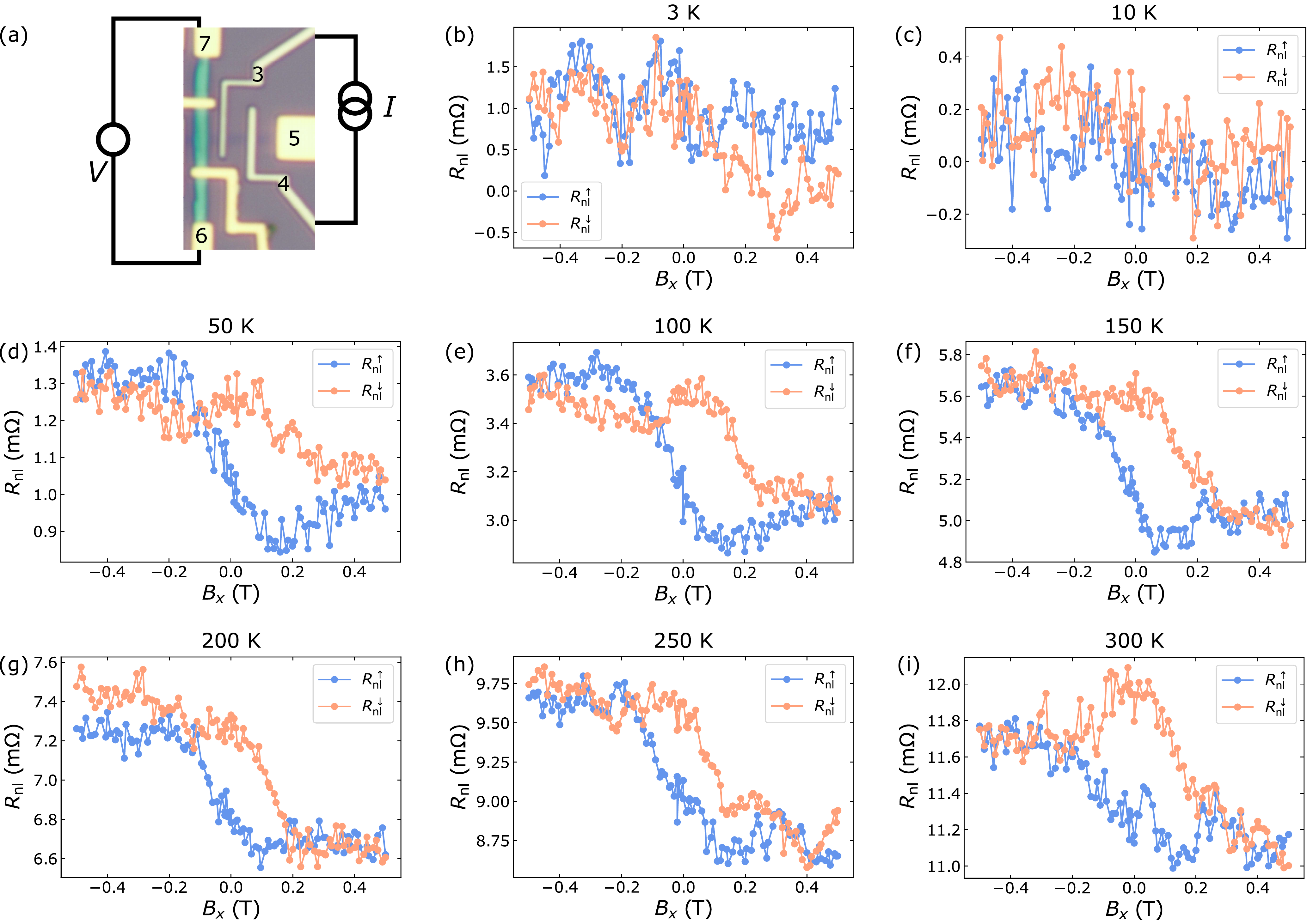}
	\caption{Spin-to-charge conversion measurements at \Vbg{}=50~V. (a) Measurement configuration. (b-i) Nonlocal SCC signal vs $B_x$ with the spin injector magnetization pointing along $+y\equiv\uparrow$ and $-y\equiv\downarrow$ and for $T$ from 3 to 300~K.}
	\label{FigureSCC1}
\end{figure}

\begin{figure}[h]
	\centering
		\includegraphics[width=\textwidth]{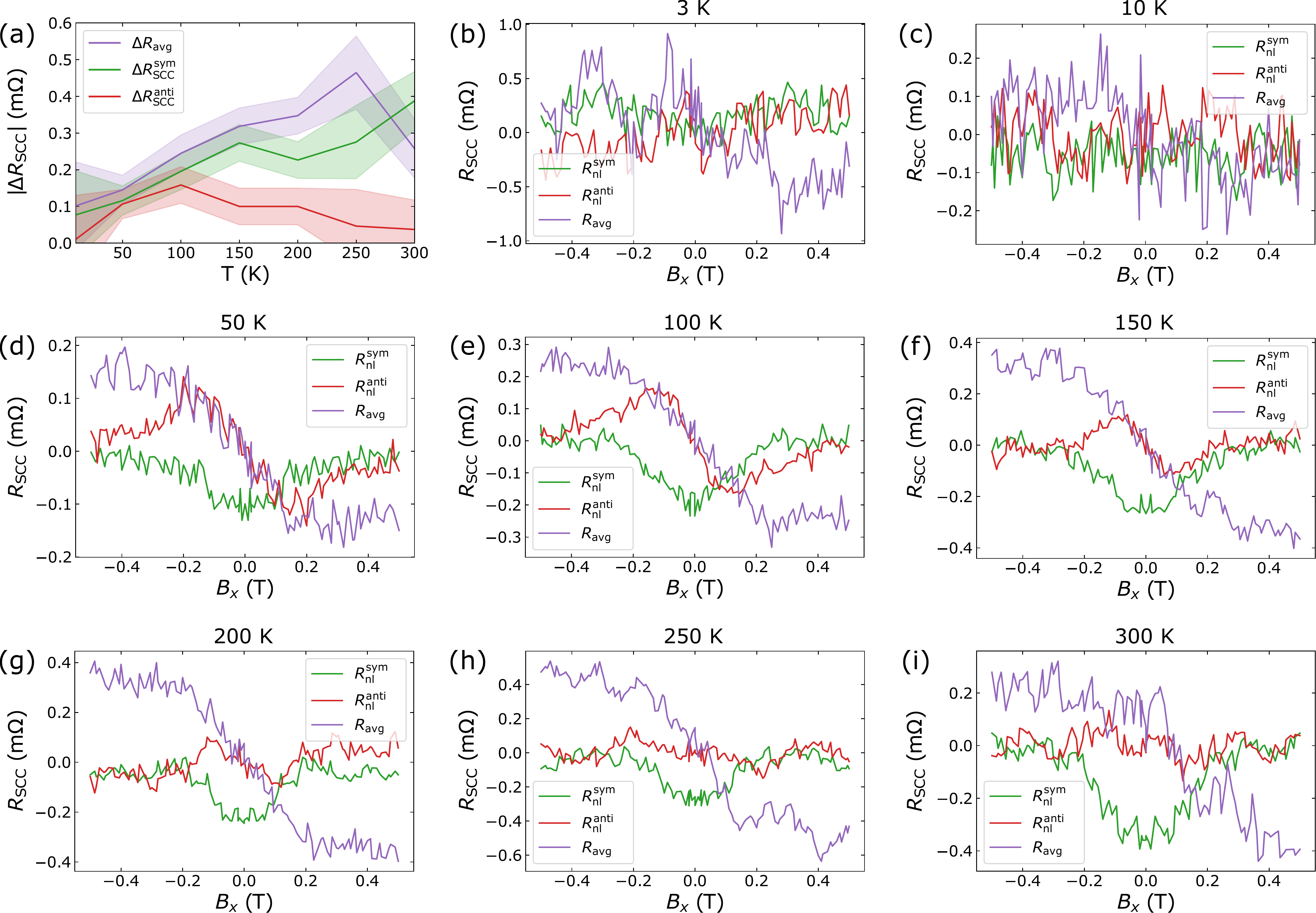}
	\caption{Spin-to-charge conversion components at \Vbg{}$=50$~V extracted from Fig.~\ref{FigureSCC1}b-i. (a) Maximum signal ($|\Delta \Rscc{}|$) vs $T$. (b-i) SCC signal as a function of $B_x$. $R_\mathrm{SCC}^\uparrow$ and $R_\mathrm{SCC}^\downarrow$ have been obtained by subtracting $R_\mathrm{nl}^\uparrow$ from $R_\mathrm{nl}^\downarrow$ and \Ravg{} by averaging. For visualization purposes, an offset has been subtracted from each \Ravg{} curve that is determined by its average value.}
	\label{FigureSCC2}
\end{figure}

The spin-to-charge (SCC) conversion experiments shown in Figs.~2 and 3 of the main manuscript are performed at temperatures from 3~K up to room temperature and \Vbg{}$=50$~V. The raw data, measured in the geometry of Fig.~\ref{FigureSCC1}a are shown in Fig.~\ref{FigureSCC1}b-i.
At 3~K, the NbSe$_2$ is superconducting and the signal offset is of about 1~m$\Omega$ and \Rup{} is very similar to \Rdown{}, which is a bit smaller for $B_x>0$. Because this signal is comparable to the noise level, we cannot confirm we have SCC in the superconducting regime.
At 10~K, the background is lower than 0.2~m$\Omega$ and the signal is below the noise level, which is about 0.1~m$\Omega$. For $T\geq 50$~K, \Rup{} and \Rdown{} show omnidirectional SCC.
In particular, the crossing of \Rup{} and \Rdown{} at negative $B_x$ indicates that $\Rscc{}=(\Rup{}-\Rdown{})/2$ contains a symmetric and an antisymmetric component. These features arise from $y$- and $z$-SCC (Figs.~1b and 1c of the main manuscript). Additionally, the average between both curves ($\Ravg{}=(\Rup{}+\Rdown{})/2$) saturates at a higher value for negative than positive $B_x$, showing that $x$-spins are also converted (Fig.~1d of the main manuscript). 
To quantify the magnitude of each SCC component, we have calculated \Rscc{} and \Ravg{}. Additionally, to distinguish between the $y$- and $z$-SCC components, we have symmetrized (antisymmetrized) the \Rscc{} data and obtained $\Rscc^\mathrm{sym(anti)}$. For this purpose, we have used
\begin{equation}
\Rscc^\mathrm{sym(anti)}=\frac{\Rscc{}(B_x)+(-)\Rscc{}(-B_x)}{2}.
\end{equation}   
Finally, to compare between the different signals, in Fig.~2 of the main manuscript we define the signal amplitudes $\Delta$\Ravg{}, $\Delta$\Rsccsym{} and $\Delta$\Rsccanti{} as the semi-difference between the maximum and minimum signal vs $B_x$.
The result of these operations (Fig.~\ref{FigureSCC2}a and Fig.~3a of the main manuscript) is that the in-plane SCC components increase with increasing $T$, in contrast with the $z$-SCC component. The $B_x$-dependent signals are shown for all $T$ in Fig.~\ref{FigureSCC2}b-i for completeness.

\section{Reciprocity}
\begin{figure}[h]
	\centering
		\includegraphics[width=0.7\textwidth]{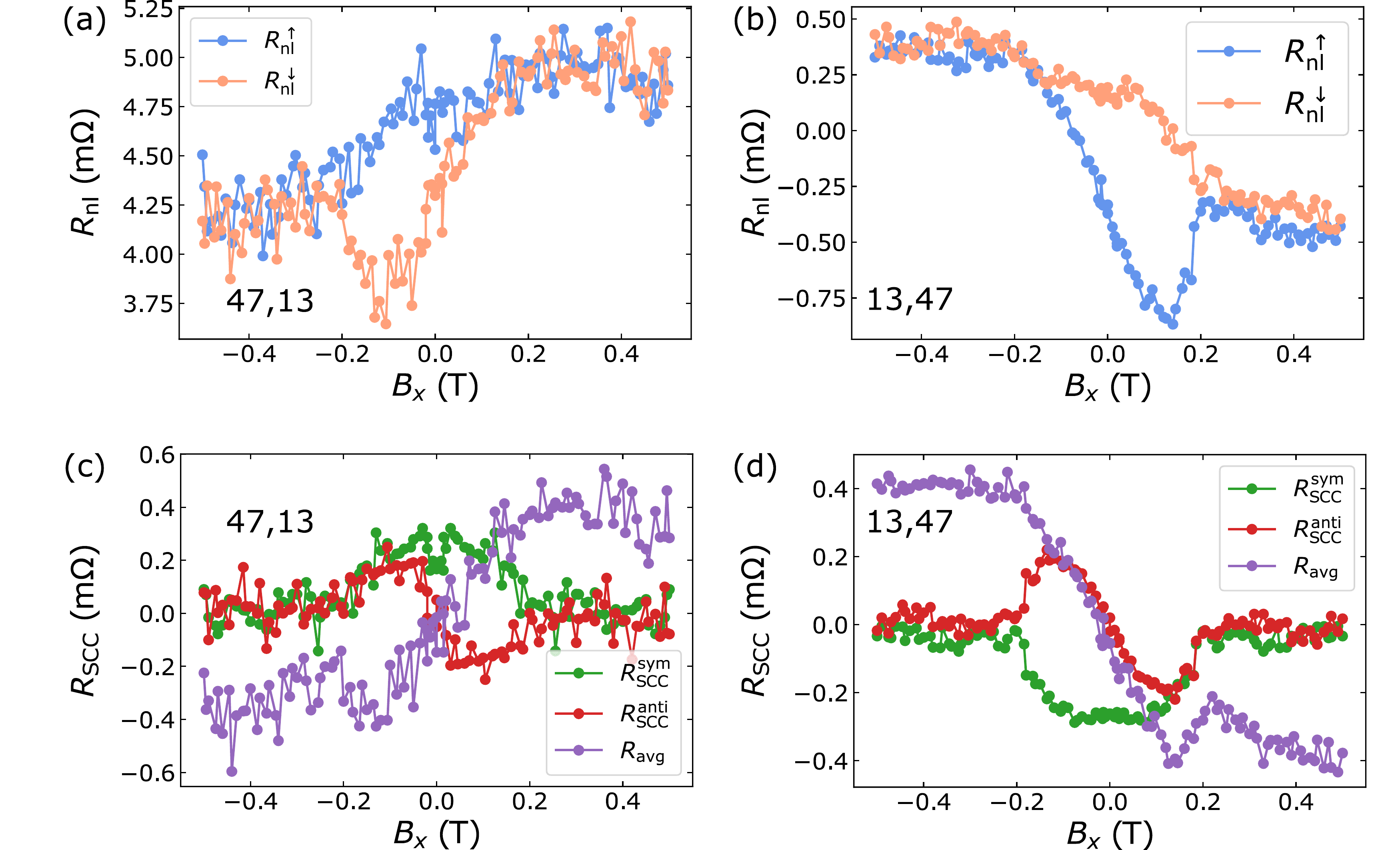}
	\caption{Spin-to-charge conversion [(a) and (b)] and corresponding SCC components [(c) and (d)] in the two reciprocal configurations at 100~K and \Vbg{}=70~V. In panels (a) and (c), $I$ is applied between contacts 4 and 7 and $V$ measured between 1 and 3. In panels (b) and (d) , $I$ is applied between contacts 1 and 3 and $V$ measured between 4 and 7. For visualization purposes, an offset has been subtracted from each \Ravg{} curve that is determined by its average value.}
	\label{FigureReciprocity}
\end{figure}
To confirm that our multiterminal measurements are in the linear response regime, we have measured the reciprocity of our signals comparing $R_{ij,kl}$ with $R_{kl,ij}$, where $i,\,j,\,k,$ and $l$ represent the $I_+,\,I_-,\,V_+,$ and $V_-$ contacts, respectively. The results from such measurement are shown in Figs.~\ref{FigureReciprocity}a and \ref{FigureReciprocity}b and each SCC component is shown separately in Figs.~\ref{FigureReciprocity}c and \ref{FigureReciprocity}d. The most important observation is that all the components keep a very similar magnitude. When looking at the signs, we see that the $x$- and $y$-SCC components reverse sign between both configurations whereas the $z$-SCC component does not.

The reciprocity theorem states that \cite{buttiker1986,buttiker1988} \begin{equation}
R_{ij,kl}(B, M)=R_{kl,ij}(-B, -M),
\label{EquationRecip1}
\end{equation}
where $M=$~$\uparrow$ or $\downarrow$ is the contact magnetization.
From this expression, one would naively expect that the three SCC components must reverse sign when swapping voltage and current terminals. However, we observe that our data, plotted in Fig.~\ref{FigureReciprocity}, shows that \Rsccanti{} does not change sign. This lack of a sign reversal can be understood by applying Equation~\ref{EquationRecip1} to our measurements:
\begin{equation}
R_\mathrm{nl}^{47,13}(B_x,\uparrow(\downarrow))=R_\mathrm{nl}^{13,47}(-B_x,\downarrow(\uparrow)).
\label{EquationRecip2}
\end{equation}
We observe that the data in Figs.~\ref{FigureReciprocity}a and \ref{FigureReciprocity}b agrees well with Equation~\ref{EquationRecip2} with the exception of a 4.5~m$\Omega$ background which we attribute to the different resistances of the graphene and NbSe$_2$ arms. To understand how reversing the voltage and current probes affect the different SCC components, we convert Equation~\ref{EquationRecip2} into
\begin{equation}
R_\mathrm{nl}^{47,13}(B_x,\uparrow)-R_\mathrm{nl}^{47,13}(B_x,\downarrow)=R_\mathrm{nl}^{13,47}(-B_x,\downarrow)-R_\mathrm{nl}^{13,47}(-B_x,\uparrow)
\label{EquationRecip3}
\end{equation}
by subtracting the two relations Equation~\ref{EquationRecip2} contains.

Next, we define $R_\mathrm{SCC}^{47,13(13,47)}(B_x)=(R_\mathrm{nl}^{47,13(13,47)}(B_x,\uparrow)-R_\mathrm{nl}^{47,13(13,47)}(B_x,\downarrow))/2$. Using this definition Equation~\ref{EquationRecip3} becomes
\begin{equation}
R_\mathrm{SCC}^{47,13}(B_x)=-R_\mathrm{SCC}^{13,47}(-B_x).
\label{EquationRecip4}
\end{equation}
The $y$-SCC component ($R_\mathrm{SCC}^{47,13\mathrm{sym}}$) is symmetric with respect to $B_x$. This means that it fulfills $R_{\mathrm{SCC}}^{47,13\mathrm{sym}}(B_x)=R_{\mathrm{SCC}}^{47,13\mathrm{sym}}(-B_x)$. Combining this expression with Equation~\ref{EquationRecip4}, we obtain that
\begin{equation}
R_{\mathrm{SCC}}^{47,13\mathrm{sym}}(B_x)=-R_{\mathrm{SCC}}^{13,47\mathrm{sym}}(B_x)
\end{equation} 
and the $y$-SCC component changes sign.

In contrast, the $z$-SCC component ($R_{\mathrm{SCC}}^{47,13\mathrm{anti}}$) is antisymmetric with $B_x$. This means that $ R_{\mathrm{SCC}}^{47,13\mathrm{anti}}(B_x)=-R_{\mathrm{SCC}}^{47,13\mathrm{anti}}(-B_x)$ and Equation~\ref{EquationRecip4} yields 
\begin{equation}
R_{\mathrm{SCC}}^{47,13\mathrm{anti}}(B_x)=R_{\mathrm{SCC}}^{13,47\mathrm{anti}}(B_x)
\end{equation}
which means that the $z$-SCC signal does not change sign.

Finally, the $x$-SCC component does not depend on the initial contact magnetization. To analyze this component separately we have used the following definition: $R_{\mathrm{avg}}^{47,13(13,47)}(B_x)=(R_\mathrm{nl}^{47,13(13,47)}(B_x,\uparrow)+R_\mathrm{nl}^{47,13(13,47)}(B_x,\downarrow))/2.$
Now we use Equation~\ref{EquationRecip2} and obtain
\begin{equation}
R_\mathrm{nl}^{47,13}(B_x,\uparrow)+R_\mathrm{nl}^{47,13}(B_x,\downarrow)=R_\mathrm{nl}^{13,47}(-B_x,\downarrow)+R_\mathrm{nl}^{13,47}(-B_x,\uparrow).
\end{equation}
That, using the definition of $R_{\mathrm{avg}}^{47,13(13,47)}(B_x)$, turns into:
\begin{equation}
R_{\mathrm{avg}}^{47,13}(B_x)=R_{\mathrm{avg}}^{13,47}(-B_x).
\label{EquationRecip5}
\end{equation}

 Because the $x$-SCC component is antisymmetric with $B_x$, $R_\mathrm{avg}^{47,13}(B_x)=-R_\mathrm{avg}^{47,13}(-B_x),$ and Equation~\ref{EquationRecip5} leads to $$R_\mathrm{avg}^{47,13}(B_x)=-R_\mathrm{avg}^{13,47}(B_x),$$ which implies that the $x$-SCC component changes sign.

 In summary, all SCC components properly fulfill reciprocity, which implies a sign reversal of the $y$- and $x$-SCC and no sign reversal of the $z$-SCC component.
 \begin{figure}[h]
	\centering
		\includegraphics[width=0.8\textwidth]{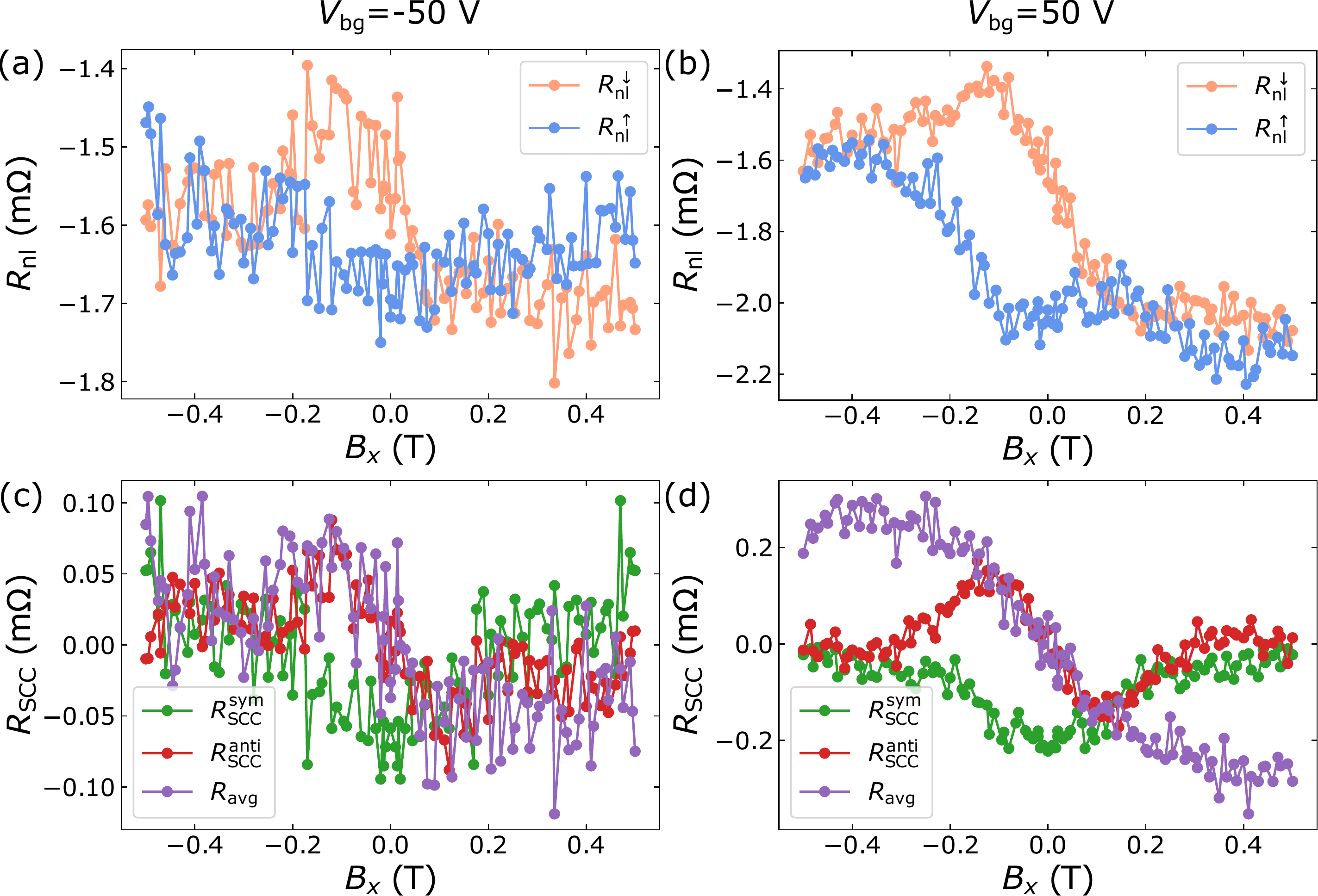}
	\caption{SCC signal as a function of $B_x$  at $T=100~K$, \Vbg{}$=-50$~V (a) and \Vbg{}$=50$~V (b). (c) and (d) SCC components extracted from panels a and b, respectively. For visualization purposes, an offset has been subtracted from each \Ravg{} curve that is determined by its average value.}
	\label{FigureSCCPM50V}
\end{figure}
 \section{Spin-to-charge conversion at $\Vbg{}=\pm50$~V}
Here we show the SCC conversion data obtained at $T=100$~K and at \Vbg{}$=\pm50$~V that shines some light on the gate tuneability of the effect.
The results are shown in Fig.~\ref{FigureSCCPM50V} and demonstrate that omnidirectional SCC is still present at \Vbg{}$= -50$~V, even though its amplitude is significantly smaller than at \Vbg{}$= 50$~V. At \Vbg{}$= 0$ we could not measure any signal below a noise level of 0.5~m$\Omega$.

\section{Determination of the interface resistance}
\begin{figure}[h]
	\centering
		\includegraphics[width=0.9\textwidth]{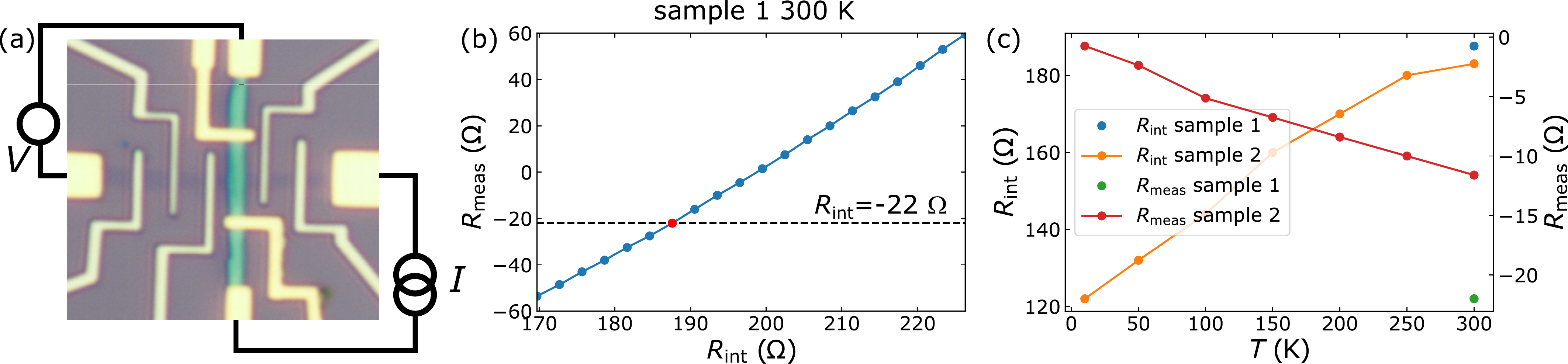}
	\caption{Determination of the interface resistance. (a) Measurement geometry used to obtain $\Rmeas{}=V/I$. (b) Relation between \Rint{} and \Rmeas{} from finite element calculations. The intersection between the horizontal dashed line (measured \Rmeas{}) and the finite-element \Rmeas{}(\Rint{}), determines the actual \Rint{} value (red dot). (c) Summary of \Rint{} and \Rmeas{} of samples 1 and 2 at different temperatures.}
	\label{FigureRint}
\end{figure}

To determine the interface resistance (\Rint{}) between the bilayer graphene and NbSe$_2$ flakes, we have used the measurement geometry shown in Fig.~\ref{FigureRint}a. Because \Rint{} is smaller than \Rsq{} of the graphene channel, the measured resistance (\Rmeas{}) is negative. Thus, to obtain the correct \Rint{} value, we used finite element calculations as in Ref.~\cite{ontoso2019}. The result of such operation is shown in Fig.~\ref{FigureRint}. Because the electrical contacts in sample~1 broke during the measurements due to the large \Vbg{} required to measure SCC, we could not obtain its \Rint{} at all the measurement temperatures. However, since sample~2 showed a very similar \Rint{} at 300~K and \Rint{} could be measured for all $T$ values, we have used the low $T$ \Rint{} values from sample~2 (red dots in Fig.~\ref{FigureRint}c) to perform the quantitative analysis. Note that, even though the value of \Rmeas{} is not the same for both samples, the different device dimensions lead to very similar \Rint{} in both cases, in agreement with the fact that both samples were prepared under the same conditions in an inert atmosphere.
\section{Estimation of the spin Hall angles in $\mathrm{NbSe}_2$}
To determine the spin Hall angles assuming that the SCC occurs in the NbSe$_2$ flake, we have solved the Bloch equations:

\begin{equation}
D_s\frac{d^2\vec{\mu}}{dx^2}-\frac{\vec{\mu}}{\tau_s}-\vec{\omega}\times\vec{\mu}=0
\label{BlochEq}
\end{equation}
Here $\vec{\mu}=(\mu_{sx}, \mu_{sy},\mu_{sz})$ is the spin accumulation, $D_s$
the spin diffusion coefficient and $\tau_s$
the spin lifetime. $\vec{\omega}=g\mu_B \vec{B}$ where $\vec{\omega}$ is the Larmor frequency, $g = 2$ the Landé factor, $\mu_B$ the Bohr
magneton, and $\vec{B}=(B_x, B_y,B_z)$ the applied magnetic field.

The first term in Equation~\ref{BlochEq} accounts for spin diffusion, the second one for spin relaxation and the last one for spin precession around a magnetic field $\vec{B}$.

\begin{figure}[h]
	\centering
		\includegraphics[width=0.7\textwidth]{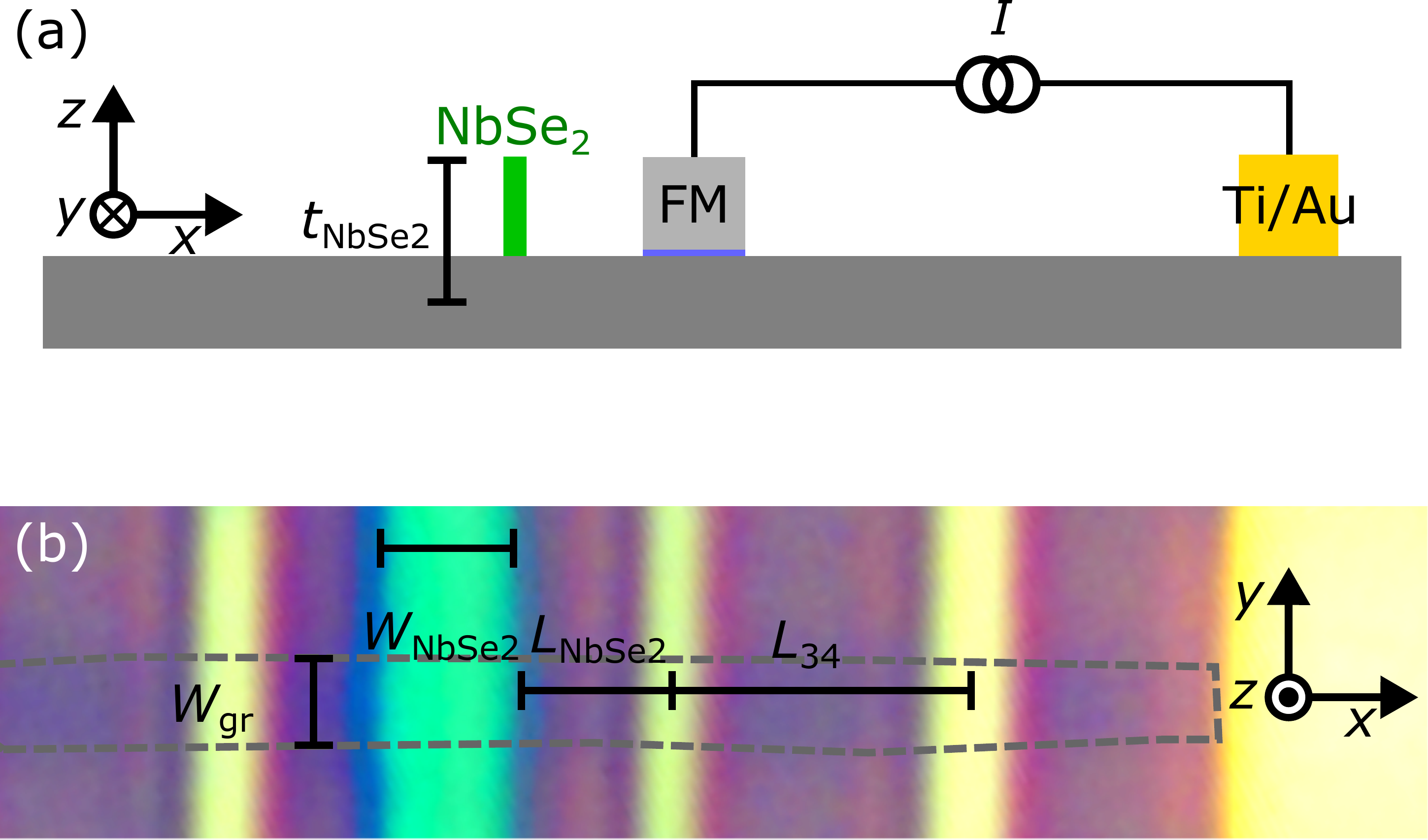}
	\caption{Geometry of the simulated device (a). The BLG (dark grey) is contacted by two ferromagnets (FM) with a tunnel barrier and the NbSe$_2$ flake (green) that is assumed to be placed at the edge that is closest to the FM spin injector where $I$ is applied. (b) Device image with the relevant dimensions. The BLG edges are indicated as a grey dashed line. }
	\label{FigureSketchModel}
\end{figure}

The modelled geometry is shown in Fig.~\ref{FigureSketchModel} and takes into account spin precession in the BLG but not in the NbSe$_2$, which has a much shorter spin lifetime. Backflow at the spin injector is also taken into account \cite{maassen2012}. Our model, which is described in detail in Ref.~\cite{ontoso2019}, introduces the effect of \Rint{} between the BLG and NbSe$_2$ flakes. The relevant parameters used for the model are shown in Table~\ref{TableS1}.

\begin{figure}[h]
	\centering
		\includegraphics[width=0.5\textwidth]{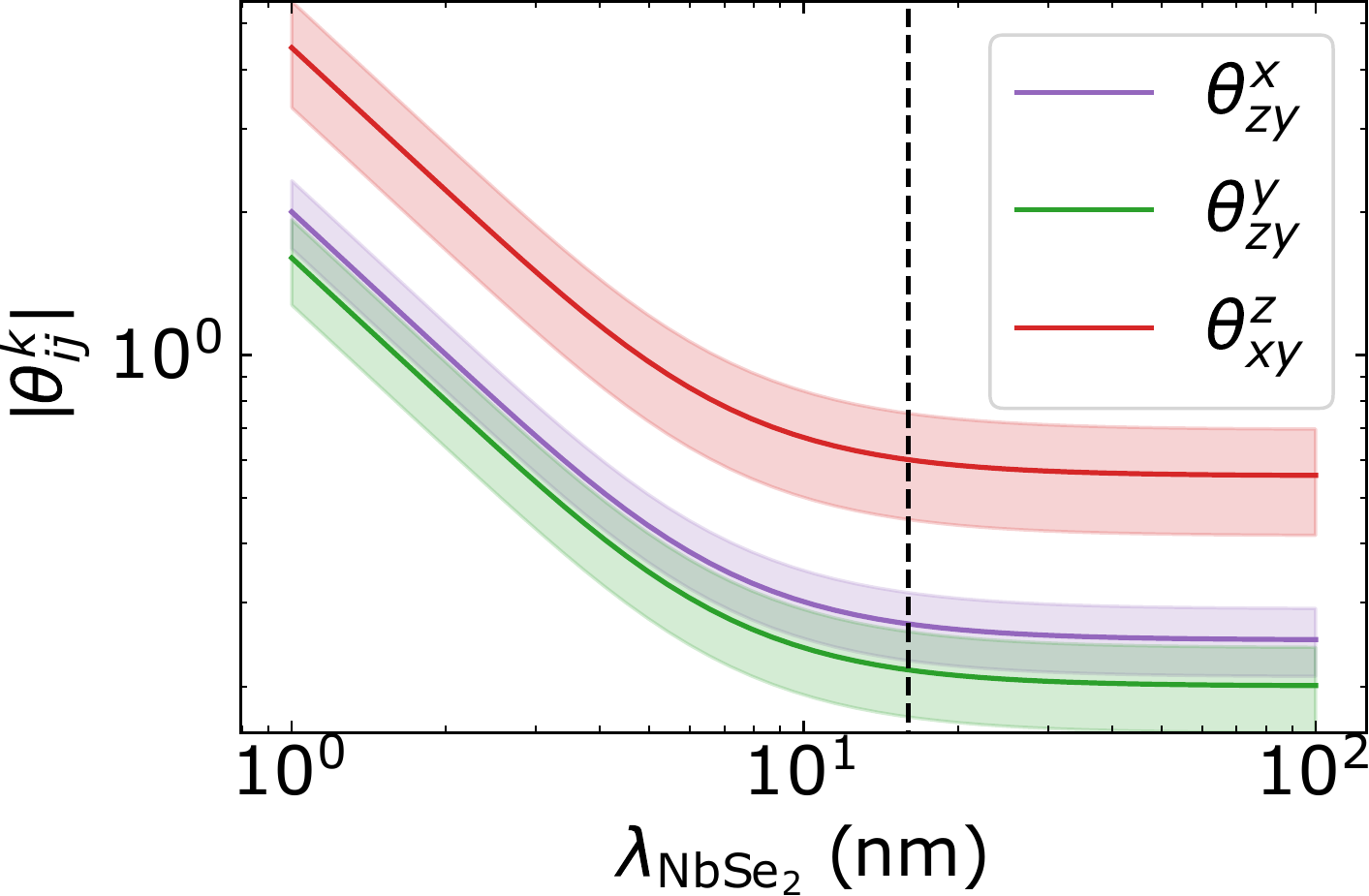}
	\caption{Spin Hall angle as a function of the spin relaxation length in the NbSe$_2$ flake.}
	\label{FigureSHAvsLambda}
\end{figure}

To determine the lower bound of the spin Hall angle ($\SHAijk{}$), where $i$, $j$, and $k$ are the directions of \js{}, \jc{} and $s$, respectively, we assume that the SCC occurs at the edge of the NbSe$_2$ that is closer to the spin injector and the width of the NbSe$_2$-covered region is smaller than the spin diffusion length of the proximitized graphene. Since the spin signal measured across the NbSe$_2$-covered region is smaller than the noise level, we cannot determine the spin relaxation length of the NbSe$_2$ (\LsNbSe2{}). Therefore, we determine \SHAijk{} as a function of \LsNbSe2{} (Fig.~\ref{FigureSHAvsLambda}). Since \SHAijk{} decreases when increasing \LsNbSe2{} until it saturates for \LsNbSe2{} larger than the NbSe$_2$ thickness, that is 16~nm in our device, we report the lower bound of the \SHAijk{} [\SHAijkmin{}] obtained assuming \LsNbSe2{}$=100$~nm. Because we do not know the actual sign of the spin polarization of the Co/TiO$_x$ contacts, we have extracted the absolute value of \SHAijkmin{} [\SHAijkminabs{}].
\begin{table}[t]
    \caption{Geometrical device parameters, contact resistance of the ferromagnetic spin injector ($R_\mathrm{c}^\mathrm{FM}$), and spin transport parameters used for the \SHAijk{} quantification summarizing Table~\ref{TableRefHanle}.}
        \begin{ruledtabular}
        \begin{tabular}{c c c c c c c c c c}
            \renewcommand{\arraystretch}{2}
            $W_\mathrm{gr}^{34}$ & $L_{34}$&$W_\mathrm{gr}$&$L_\mathrm{NbSe_2}$& $W_\mathrm{NbSe_2}$&$t_\mathrm{NbSe_2}$&$R_\mathrm{c}^\mathrm{FM}$&$\tau_s$&$D_s$&$P$\\
			($\mu$m)	&($\mu$m)&($\mu$m)	&($\mu$m)&($\mu$m)&(nm)&(k$\Omega$)&(ps)&(m$^2$/s)&\\
            \hline
             1.0 & 2.0 &0.80&1.0&0.9&16&6.3&20&0.01&0.035\\
    \end{tabular}
    \end{ruledtabular}
    \label{TableS1}
\end{table}

\section{Spin precession with out-of-plane magnetic field in sample~2}
\begin{figure}[h]
	\centering
		\includegraphics[width=0.7\textwidth]{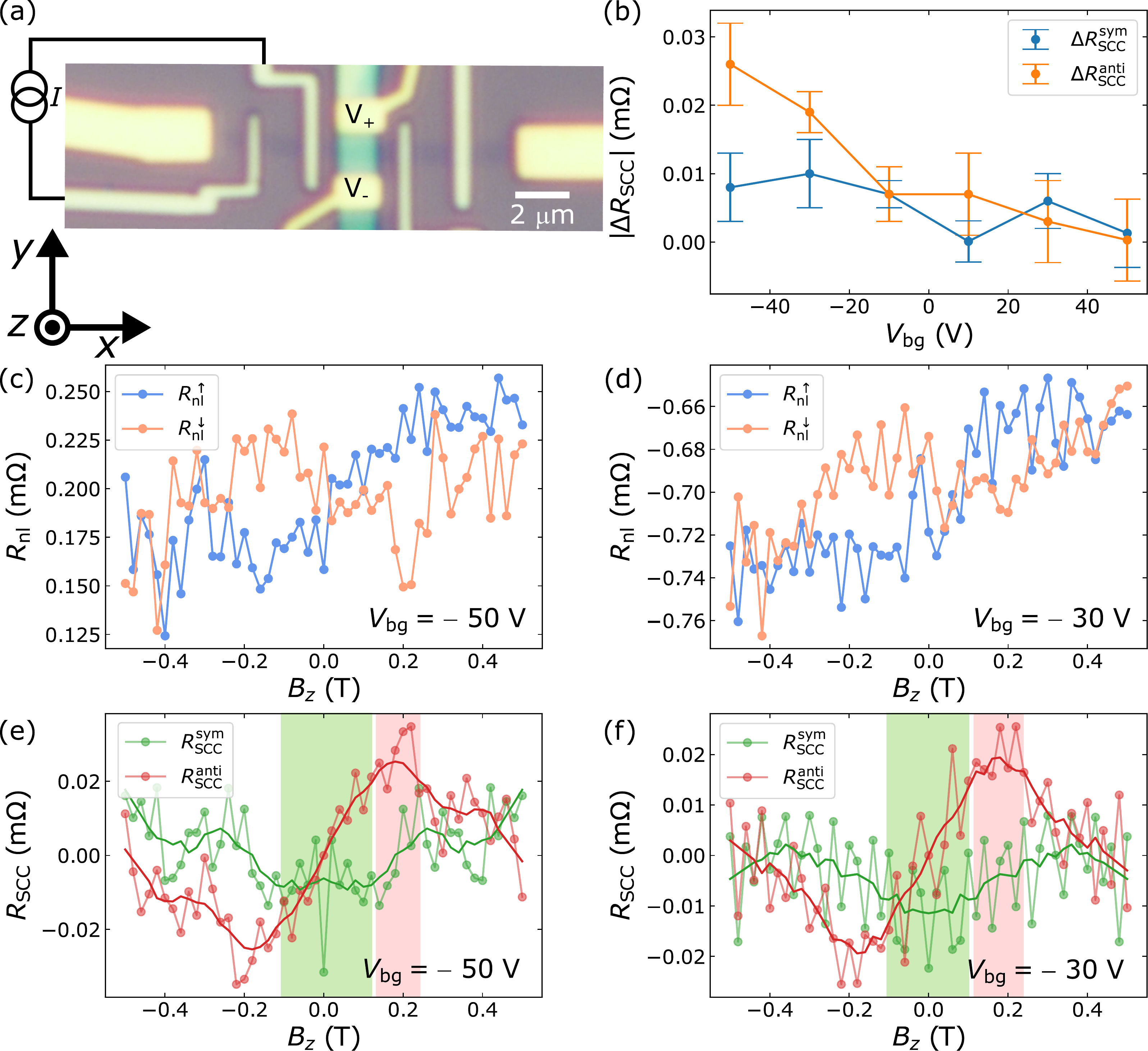}
	\caption{Spin-to-charge conversion as a function of $B_z$ in sample 2 at 100~K. (a) Optical microscope image of sample 2 with the measurement circuit. (b) \Vbg{}-dependence of $\Delta R_\mathrm{SCC}$. (c) and (d) SCC at \Vbg{}$=-50$ and $-30$~V, respectively. (e) and (f) Symmetric ($y$-SCC) and antisymmetric ($x$-SCC) components of \Rscc{}$=(\Rup{}-\Rdown{})/2$ at \Vbg{}$=-50$ and $-30$~V, respectively. The green and red rectangles highlight the areas that have been used to determine the amplitude and error, that is two standard deviations from the mean for the symmetric and antisymmetric component, respectively. The results are shown in panel b.}
	\label{FigureSample2}
\end{figure}
We have also performed SCC experiments in sample 2. 
In this case, to demonstrate that the $x$-SCC component is indeed due to spin transport, we have applied an out-of-plane magnetic field ($B_z$). $B_z$ induces spin precession in the $x-y$ plane. As a consequence, $x$-SCC appears as the antisymmetric component in \Rscc{} and $y$-SCC as the symmetric one. The results from sample 2 are summarized in Fig.~\ref{FigureSample2}. There, one can identify both a symmetric and antisymmetric component in \Rscc{}, which show that both $x$- and $y$-SCC components are present in sample 2. This observation confirms that the $x$-SCC component is due to SCC and that both in-plane SCC components are achieved in multiple samples.  Another observation we make is that the SCC signal is one order of magnitude smaller than in sample 1, making its observation very challenging. As a consequence, we only observed SCC at 100~K in sample 2. Additionally, for sample 2, we only observed SCC at \Vbg{}$=-50$ and $-30$~V, in contrast with sample 1, where we observed it at positive \Vbg{}. We attribute this observation, that is summarized in Fig.~\ref{FigureSample2}b, to a different doping of the BLG flake that gives rise to better spin transport for holes than electrons, in contrast with sample 1.
\section{Determination of the rotation angle between bilayer graphene and NbSe$_2$}
\begin{figure}[h]
	\centering
		\includegraphics[width=\textwidth]{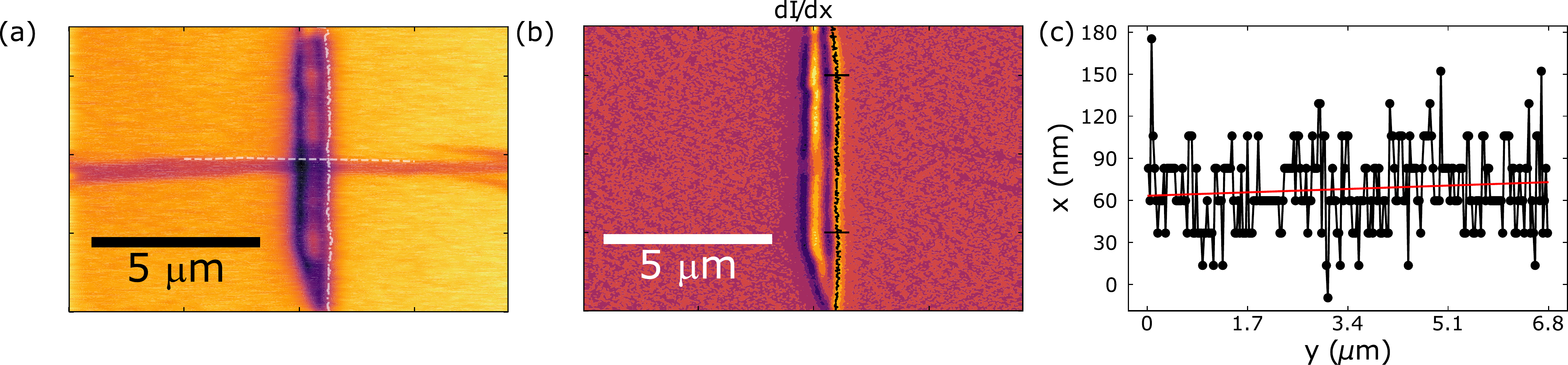}
	\caption{Determination of the flake rotation. (a) Optical microscope image of sample 1 before contact deposition with the estimated edges of the NbSe$_2$ (vertical) and BLG (horizontal) as dashed lines. (b) Derivative vs $x$ of the intensity of the image in panel a used to determine the edge of the NbSe$_2$ (dashed line). The horizontal lines limit the range that is used. (c) Linear fit of the points obtained from panel b to obtain the angle between the NbSe$_2$ edge and the $x$-axis.}
	\label{FigureRotation}
\end{figure}

To determine the rotation angle between the BLG and NbSe$_2$ crystal mirrors, we have assumed that the flakes cleave along a crystallographic axes \cite{you2008}. In this case, the alignment between the mirrors can be found by analyzing the optical microscope images of sample~1 and finding the rotation angle between the straight edges. To obtain the edge of the NbSe$_2$ flake in the most accurate way, we have taken the derivative of the image obtained from averaging the red, green and blue channels of the image (Figs.~\ref{FigureRotation}a and \ref{FigureRotation}b) and traced its maximum for each value of $y$. Next, we have fit the extracted data to a line (Fig.~\ref{FigureRotation}c, $x=A+By$ where $x$ is the horizontal and $y$ the vertical coordinate in the image) to obtain its rotation angle with respect to the image using $\tan(\alpha_\mathrm{NbSe_2})=B$. The confidence interval of $\alpha_\mathrm{NbSe_2}$ has been obtained from the confidence interval of the slope obtained from the fit and assuming that the uncertainty in the determination of the data is its standard deviation from the average value (note that this should lead to a small overestimation due to the small slope of the $x(y)$ line). The result from such analysis is $\alpha_\mathrm{NbSe_2}=89.91^\circ\pm 0.06^\circ$.

To determine the rotation angle of the graphene flake with respect to the image, because its contrast is too small and taking the derivative results in a noisy image, we have determined the edges by visual inspection selecting 7 points that we have plotted as a dashed line in Fig.~\ref{FigureRotation}a. From the slope of this line, we have determined the angle of the graphene flake ($\alpha_\mathrm{BLG}$), that is $\alpha_\mathrm{BLG}=-0.9^\circ\pm0.6^\circ$. The confidence interval has been obtained assuming that the uncertainty in $y$ is 0.05~$\mu$m and $\delta\alpha_\mathrm{BLG}=\arctan(\sqrt{2}\times0.05\,\mu$m$/6.8\,\mu$m), where 6.8 $\mu$m is the length of the dashed line. The factor $\sqrt{2}$ comes from the assumption that the uncertainty in the determination of the different points along the edge is not correlated. The value obtained from a linear fit to all the points coincides with the result extracted from the two extreme values.

Finally, the rotation angle between both flakes has been obtained as $\alpha_\mathrm{NbSe_2}-\alpha_\mathrm{BLG}=89^\circ\pm0.6^\circ$. 

Similar analysis performed for sample 2 yields a twist angle of $92.7^\circ\pm0.8^\circ$.  

Note that these values are just an estimate and are only valid if the selected edges correspond to crystal mirrors.
\section{Symmetry considerations}
Bulk NbSe$_2$ has space group P6$_3$/mmc (194) with corresponding point group 6/mmm (D$_{6h}$) \cite{lide2004}. Hence, the spin Hall effect (SHE) has only the conventional components, that are given by
\begin{equation}
(J_i)^k=\sigma_{ij}^k E_j
\end{equation}
with $i\neq j \neq k$, where $(J_i)^k$ is the spin current density propagating along direction $i$ and polarized along $k$ \cite{culcer2007,wimmer2015, seemann2015}. $\sigma_{ij}^k$ is the spin Hall conductivity that relates the charge current density $E_j$ along $j$ with the spin current density $(J_i)^k$ via SHE.

In bulk NbSe$_2$ there is no Edelstein effect (EE) due to inversion symmetry.

Freestanding monolayer NbSe$_2$ has point group $D_{3h}$. In this case the SHE still has only conventional components, and the EE is still zero, even though it does not have inversion symmetry.

The surface of bulk NbSe$_2$ (or monolayer NbSe$_2$ on a substrate) has point group $C_{3v}$. The SHE has an extra component $\sigma_{xx}^x=-\sigma_{xy}^y=-\sigma_{yx}^y=-\sigma_{yy}^x$ which is not Hall-like (is symmetric in $i\rightarrow j$) \cite{culcer2007,wimmer2015, seemann2015,roy2021}.
The EE has one term $M_i=\nu_{ij}E_j$ with $\nu_{xy}=-\nu_{yx}$ and the other components are zero. Here, $M_i$ is the induced magnetization, $E_j$ the applied electric field, and $\nu_{ij}$ the EE tensor converting the electric field into an induced magnetization.
For these components, the mirror in $C_{3v}$ is $x\rightarrow -x$.

The unconventional $y$-SCC component that we want to characterize has $\vec{j_c}\sim y$ and $\vec{s}\sim y$. This means that the unexpected SCC component can be due to an unconventional EE component ($\nu_{yy}$) or from the SHE component $\sigma_{yz}^y$. There is an additional component which is allowed, that is $\sigma_{xy}^y$. However, because the signal does not change sign upon reversing the spin diffusion direction along $\pm x$, the latter is not allowed.
None of the two possible sources would be allowed even in the less restrictive scenario, that is the surface of bulk NbSe$_2$, thus, we have to break extra symmetries in our device to explain the measured signal.

Bulk NbSe$_2$ with shear strain would have broken $C_3$, $C_{2y}$ and $M_x$ but preserved $M_z$ and inversion. $\sigma_{yz}^y$ is allowed in this case, but not $\nu_{yy}$ because of inversion symmetry \cite{seemann2015}.

The surface of NbSe$_2$ with shear strain has no symmetry left, hence, everything is allowed.

If the mirrors of graphene's symmetry group and NbSe$_2$ are not aligned, the interface between NbSe$_2$ and graphene does not have any mirror and $\nu_{yy}$ is allowed without the need for shear strain. Where a Nb or Se atom lies on top of a C atom or a hexagon centre this results in a C$_3$ point group. As explained in the main manuscript, we believe this is the most likely scenario in our devices \cite{li2019, david2019}.
In contrast, when the twist angle is 0 or a multiple of 30 degrees, the mirrors are aligned, $\nu_{yy}$ is not allowed and, where a Nb or Se atom lies on top of a C atom or a hexagon centre, the heterostructure belongs to the C$_{3v}$ point group.
\FloatBarrier
\end{document}